\documentstyle[12pt,epsfig]{article}

\newlength{\dinwidth}
\newlength{\dinmargin}
\def\lapproxeq{\lower .7ex\hbox{$\;\stackrel{\textstyle
<}{\sim}\;$}}
\def\gapproxeq{\lower .7ex\hbox{$\;\stackrel{\textstyle
>}{\sim}\;$}}
\def\gtrsim{ \;\raisebox{-.7ex}{$\stackrel{\textstyle
>}{\sim}$}\; }

\def\be{\begin{equation}}
\def\ee{\end{equation}}
\def\bea{\begin{eqnarray}}
\def\eea{\end{eqnarray}}

\def\pp{p\bar{p}}
\def\ra{ \rightarrow }

\def\GeV{\rm GeV}
\def\TeV{\rm TeV}
\def\a{{\alpha}_S}
\def\ol{\overline }

\begin{document}
\titlepage
\begin{flushright}
IPPP/03/45 \\
DCPT/03/90 \\
Cavendish-HEP-2003/14 \\
7 August 2003 \\

\end{flushright}

\vspace*{0.5cm}

\begin{center}
{\Large \bf Uncertainties of predictions from parton
distributions\\ \vspace{1.5ex} II:\ Theoretical errors}

\vspace*{1cm}
\textsc{A.D. Martin$^a$, R.G. Roberts$^b$, W.J. Stirling$^a$
and R.S. Thorne$^{c,}$\footnote{Royal Society University Research Fellow.}} \\

\vspace*{0.5cm} $^a$ Institute for Particle Physics Phenomenology,
University of Durham, DH1 3LE, UK \\
$^b$ Rutherford Appleton Laboratory, Chilton, Didcot, Oxon, OX11 0QX, UK \\
$^c$ Cavendish Laboratory, University of Cambridge, \\ Madingley Road,
Cambridge, CB3 0HE, UK
\end{center}

\vspace*{0.5cm}

\begin{abstract}
We study the uncertainties in parton distributions, determined in global fits to deep inelastic and related hard
scattering data, due to so-called theoretical errors. Amongst these, we include potential errors due to the change
of perturbative order (NLO $\to$ NNLO), $\ln(1/x)$ and $\ln(1-x)$ effects, absorptive corrections and higher-twist
contributions. We investigate these uncertainties both by including explicit corrections to our standard global
analysis and by examining the sensitivity to changes of the $x,Q^2,W^2$ cuts on the data that are fitted. In this
way we expose those kinematic regions where the conventional DGLAP description is inadequate. As a consequence we
obtain a set of NLO, and of NNLO, {\em conservative} partons where the data are fully consistent with DGLAP
evolution, but over a restricted kinematic domain. We also examine the potential effects of such issues as the
choice of input parameterization, heavy target corrections, assumptions about the strange quark sea and isospin
violation. Hence we are able to compare the theoretical errors with those uncertainties due to errors on the
experimental measurements, which we studied previously. We use $W$ and Higgs boson production at the Tevatron and
the LHC as explicit examples of the uncertainties arising from parton distributions. For many observables the
theoretical error is dominant, but for the cross section for $W$ production at the Tevatron both the theoretical
and experimental uncertainties are small, and hence the NNLO prediction may serve as a valuable luminosity
monitor.

\end{abstract}

\newpage

\section{Introduction}

A knowledge of the partonic structure of the proton is an essential ingredient in the analysis of hard scattering
data from $pp$ or $\pp$ or $ep$ high energy collisions. However, only the $Q^2$ (or scale) dependence of the
parton distributions can be calculated from perturbative QCD. Perturbative QCD cannot fix their absolute values.
Moreover, non-perturbative techniques are still far from being able to predict reliable magnitudes. Rather, to
determine the distributions, it is necessary to resort to global analyses of a wide range of deep inelastic and
related hard scattering data. The Bjorken $x$ dependence of the distributions is parameterized at some low scale,
and a fixed order (either LO, NLO or NNLO) DGLAP evolution performed to specify the distributions at the higher
scales where data exist. Much attention is now being devoted to obtaining reliable uncertainties on the parton
distributions obtained in this way. One obvious uncertainty is due to the systematic and statistical errors of the
data used in the global fit. We will call these the {\em experimental} errors on the parton distributions and on
the physical observables predicted from them. In fact, these so-called experimental uncertainties of the partons
have so far been the main focus of attention. They have been estimated by several groups
\cite{Botje}--\cite{MRST2002}, working within a NLO framework using a variety of different procedures. For
instance, in a previous paper \cite{MRST2002} we estimated the parton errors using a Hessian approach in which we
diagonalised the error matrix and then used the linear propagation of errors to estimate the uncertainty on a
variety of typical observables. We confirmed that this simple approach works well in practice by using a more
rigorous Lagrange multiplier method to determine the errors on the physical quantities directly. We also compared
our results with those obtained in similar analyses performed by the CTEQ
collaboration~\cite{CTEQLag,CTEQHes,CTEQ6}.

Besides the {\em experimental} errors, there are many other sources of uncertainty associated with the global
parton analysis. These are the concern of the present paper. They may loosely be called {\em theoretical} errors.
That is, we use theoretical errors to denote all uncertainties on the predicted observables other than those that
arise from the systematic and statistical errors of the data that are included in the global fit. The various
theoretical errors may be divided into four categories. Uncertainties due to (i)~the selection of data fitted,
(ii)~the truncation of the DGLAP perturbation expansion, (iii)~specific theoretical effects\footnote{Of course,
differences can also arise from alternative methods of treating the behaviour near the heavy flavour thresholds.
However, this issue is now well understood. Either fixed-flavour-number schemes or, preferably, one of the
selection of variable-flavour-number schemes can be used for a correct description~\cite{HeavyFlav}. Different
choices will result in slight differences in the extracted partons, but only minimal variation in predictions for
physical quantities.} (namely $\ln 1/x$, $\ln(1-x)$, absorptive and higher twist corrections) and (iv)~input
assumptions (such as the choice of parameterization, heavy target corrections, whether isospin violation is allowed 
and the form of the strange quark sea). These theoretical errors are discussed in 
turn in the following four sections. Then, in
Section~\ref{sec:implications}, we study the implications of the error analysis for determining the value of
$\a(M_Z^2)$ from the global fit, and for the accurate predictions of particularly relevant observables
at the Tevatron and the LHC.

\section{Uncertainties due to the selection of data fitted} \label{sec:selection}

In principle, if the DGLAP formalism is valid and the various data sets are compatible, then changing the data
that are included in the global analysis should not move the predictions outside the error bands. In practice this
is not the case. As an extreme example, consider the analysis of the H1 and BCDMS deep inelastic data for
$F_2$~\cite{H1Krakow}. The absence of the other fixed target data and, in particular, the Tevatron jet data from
the global parton fit results in a gluon which is much smaller for $x\gtrsim0.3$, and hence larger at small $x$,
from those obtained in global fits to a wider range of data. Another example occurs in the analysis of
Ref.~\cite{Giele}, where fitting to a small subset of the total data (namely H1(94)~\cite{H194},
BCDMS~\cite{BCDMS} and E665~\cite{E665} data for $F_2^p$) yields a surprisingly low value $\alpha_S = 0.112\pm
0.001$, and a very hard high $x$ down-quark distribution. Even fits of different groups, at the same perturbative order, to
basically the same data lead to unexpectedly sizable differences in partons and in predictions for observables.
For example, compare the MRST2002 and CTEQ6 predictions at NLO for the cross sections for $W$ and Higgs boson
production at hadron colliders, shown in Fig.~15 of Ref.~\cite{MRST2002}. We see that the CTEQ6 values at the
Tevatron are more than $1.5\%$ smaller for $W$ production and 8\% smaller for Higgs production; differences which
are larger than expected from an analysis based on the experimental errors of the data used in the fits.

We begin by investigating the stability of the global parton
analysis to different choices of the data cuts ($W_{\rm cut}$,
$x_{\rm cut}$, $Q^2_{\rm cut}$), defined such that data with
values of $W$, $x$ or $Q^2$ below the cut are excluded from the
global fit, with the implicit assumption that the remaining data
can be described by pure leading-twist DGLAP evolution. We find
the minimum values of the data cuts for which this stability
occurs. The kinematic variable $W$ is the invariant mass of the
system $X$ recoiling against the scattered lepton in deep
inelastic lepton--proton scattering $lp\ra lX$. It follows that
\be W^2 \simeq Q^2(1-x)/x. \label{eq:Wsquared} \ee
In the remainder of this section, all fits are performed at next-to-leading order (NLO)
unless otherwise stated. In Section~\ref{sec:change} we discuss what happens when we use
the NNLO formalism.

\subsection{Effect of the cut on $W^2$} \label{sec:effectWsquared}

In the original MRST global analyses we have fitted to data with
$W^2>10~\GeV^2$, assuming that this was sufficiently high to avoid
resonance structure, large $\ln(1-x)$ effects and associated
higher-twist corrections. However, we had no quantitative
justification for this precise choice, and, noting that it
resulted in a systematically poor fit to SLAC data with
$W^2\gtrsim 10~\GeV^2$, we subsequently raised the cut to $W^2_{\rm cut} =
12.5~\GeV^2$~\cite{MRST2001}. This provided an acceptable
description of the SLAC data. In order to make a more systematic
investigation of the stability of the fit to varying this cut, we
performed a series of global analyses with $W^2_{\rm cut}$ ranging
from 12.5 to $25~\GeV^2$. When raising the cut from 12.5 to
$15~\GeV^2$ we find $\chi^2$ to the remaining data improved by
only 4, while an increase of $W^2_{\rm cut}$ from 15 to 20 or
$25~\GeV^2$ resulted in no significant improvement. Taking these
results at face value, we conclude that $W^2_{\rm cut} =
15~\GeV^2$ is a conservative choice and that there is no reason to
eliminate even more data.

However, inspection of the description of the SLAC and BCDMS data
in this low $W^2$ region shows a lack of compatibility of the two
data sets in the region where they overlap; specifically for $W^2$
in the interval $(6,15)~\GeV^2$ for the higher values of $x$. Thus
the stability achieved at $W^2_{\rm cut} = 15~\GeV^2$ corresponds
to the SLAC data completely disappearing from the fit. Hence, while the
resulting fit describes the BCDMS data well, an extrapolation to
only slightly lower $W^2$ values gives a very poor description of
the SLAC data. This implies that the achieved stability is an
artifact of the incompatibility of the two data sets for
$F_2^{p,n}$. Indeed, when phenomenological higher-twist
contributions are introduced into the analysis they still have
significant impact for $W^2>15~\GeV^2$, see Fig.~2 of
Ref.~\cite{MRSTtwist}. This implies that a genuinely conservative
choice of cut would be $W_{\rm cut}\sim 20~\GeV^2$. However,
because a good description is obtained of the only available data
in the $(15,20)~\GeV^2$ interval we may set $W_{\rm cut} =
15~\GeV^2$ without prejudicing the analyses. Future measurements
of $F_2^p$ at HERA in this kinematic domain would clearly be
valuable.

\subsection{Effect of the choice of the cut on $x$}
\label{sec:effectx}

In Table~1 we show the values of $\chi^2$ for global analyses
performed for different values of $x_{\rm cut}$, together with the
number of data points fitted. Each column represents the $\chi^2$
values corresponding to a fit performed with a different choice of
the cut in $x$. For example, if only data above $x_{\rm cut} =
0.001$ are fitted, then the total $\chi^2 = 2119$, with a
contribution of $\chi^2 = 2055$ coming from the subset of data
with $x>0.0025$, and $\chi^2 = 2012$ from the subset with
$x>0.005$, etc.

\begin{table}[h]\begin{center}
\begin{tabular}{|c|c|c|c|c|c|c|} \hline
$x_{\rm cut}:$ & 0 & 0.0002 & 0.001 & 0.0025 & 0.005 & 0.01 \\
\hline \# data points & 2097 & 2050 & 1961 & 1898 & 1826 & 1762 \\
$\alpha_S(M_Z^2)$ & 0.1197 & 0.1200 & 0.1196 & 0.1185 &
0.1178 & 0.1180 \\ \hline $\chi^2(x>0)$ & 2267 &&&&& \\
$\chi^2(x>0.0002)$ & 2212 & 2203 &&&& \\
$\chi^2(x>0.001)$ & 2134 & 2128 & 2119 &&& \\
$\chi^2(x>0.0025)$ & 2069 & 2064 & 2055 & 2040 && \\
$\chi^2(x>0.005)$ & 2024 & 2019 & 2012 & 1993 & 1973 & \\
$\chi^2(x>0.01)$ & 1965 & 1961 & 1953 & 1934 & 1917 & 1916 \\
\hline $\Delta_i^{i+1}$ & \multicolumn{6}{| l |}{$\qquad\ \  0.19
\quad\ \ \,  0.10 \quad\ \ \;  0.24 \quad\ \ \  0.28 \quad\ \ \ \,  0.02\quad\;  $} \\
\hline
\end{tabular} \caption{\label{tab:t1}
Each column shows the $\chi^2$ values obtained from a NLO global analysis
with a different choice of $x_{\rm cut}$, together with the number
of data points fitted and the value of $\alpha_S(M_Z^2)$ obtained.
The first $\chi^2$ entry in a given column is the total $\chi^2$,
and the subsequent entries show the contributions to $\chi^2$ from
subsets of the data that are fitted. The quantity
$\Delta_i^{i+1}$, shown in the final row, is a measure of
stability to changing the choice of $x_{\rm cut}$, as explained in
the text. In these analyses we take the default cut in $Q^2$, that
is $Q^2_{\rm cut} = 2~\GeV^2$.}
\end{center}
\end{table}

To obtain a measure of the stability of the analysis to changes in the choice of $x_{\rm cut}$, we compare fits in
adjacent columns, that is with $(x_{\rm cut})_{i+1}$ and $(x_{\rm cut})_i$. In particular, it is informative to
compare the contributions to their respective $\chi^2$ values from the subset of data with $x>(x_{\rm
cut})_{i+1}$. If stability were achieved, then we would expect the difference $\Delta \chi^2$ between these two
$\chi^2$ contributions to be very small. We stress that these two $\chi^2$ contributions describe the quality of
the two fits to the {\em same} subset of the data. Thus, as we shall explain below, a measure, $\Delta_i^{i+1}$,
of the stability of the analysis is $\Delta\chi^2$ divided by the number of data points omitted when going from
the fit with $(x_{\rm cut})_i$ to the fit with $(x_{\rm cut})_{i+1}$. For example, if we raise the $x_{\rm cut}$
from 0.001 to 0.0025 then $\Delta\chi^2 = 2055 - 2040$ for the data with $x>0.0025$, and the number of data points
omitted is $1961 - 1898 = 63$. Thus the measure $\Delta_{0.001}^{0.0025} = 15/63 = 0.24$, as shown in the last row
of Table~1. If we were to start from a fit with $x_{\rm cut}$ below the value at which the theoretical framework
is expected to be valid (due to neglected $\ln 1/x$ effects, etc.), then we would expect $\Delta_i^{i+1}$ to be
significantly non-zero. As $x_{\rm cut}$ is increased, then $\Delta_i^{i+1}$ should decrease and, in the ideal
case, approach and remain near zero; indicating that we are in the stable domain where the theoretical framework
is appropriate. In this way the behaviour of $\Delta_i^{i+1}$, as $x_{\rm cut}$ is changed, acts as an indicator
of the stability of the analysis.

\begin{figure}[htbp]
\epsfig{figure=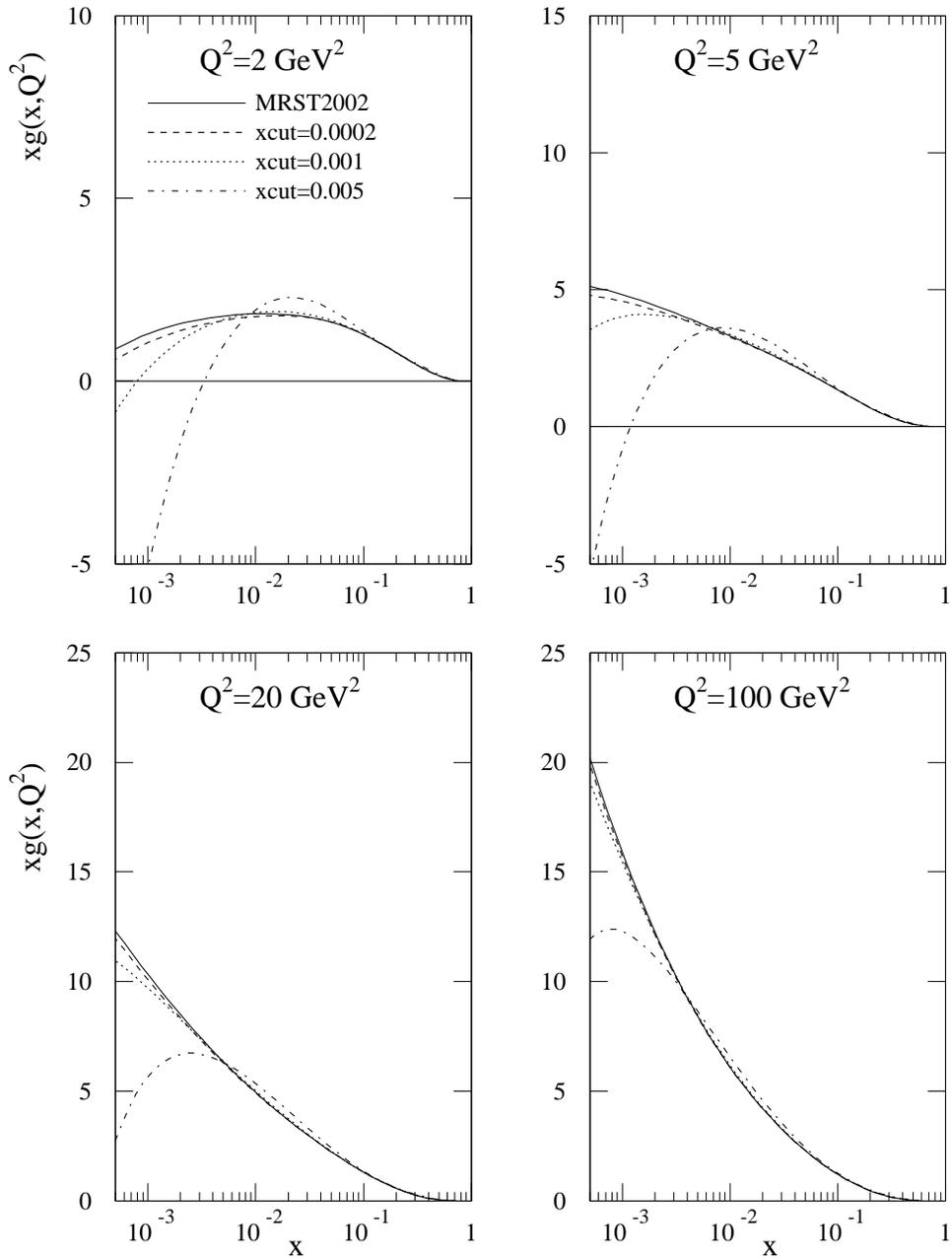,height=20cm}
 \caption{\label{fig:f1} The gluon distribution obtained in NLO global
 fits with different values of $x_{\rm cut}$, that is $x_{\rm cut}$
 taken to be 0.0002 (dashed curve), 0.001 (dotted) and 0.005 (dot-dashed),
 compared to the default set with $x_{\rm cut}=0$ (continuous
 curve).}
\end{figure}

Inspection of the values of $\Delta_i^{i+1}$ in the last row of
Table~\ref{tab:t1} shows a significant improvement in the quality
of the fit each time $x_{\rm cut}$ is raised by an amount
corresponding to the omission of about a further 70 data points,
until the final step when $x_{\rm cut}$ is increased from 0.005 to
0.01, when we see that there is no further improvement at all. In
fact, raising $x_{\rm cut}$ from 0.01 to 0.02 confirms this
stability. Indeed, $\Delta_i^{i+1}$ is greater when raising
$x_{\rm cut}$ from 0.0025 to 0.005 than in any of the previous
steps. Hence we conclude that $x\simeq 0.005$ is a safe choice of
$x_{\rm cut}$. Below this value there is a degree of
incompatibility between the data and the theoretical description.
At each step the gluon distribution extrapolated below $x_{\rm
cut}$ becomes increasingly smaller, allowing the higher $x$ gluon
to increase (and to carry more momentum), see Fig.~\ref{fig:f1}.
Because almost no momentum is carried by the gluon at very small
$x$, this redistribution of the gluon momentum becomes
increasingly possible as $x_{\rm cut}$ is raised, hence explaining
why $\Delta_i^{i+1}$ has a tendency to increase, until stability
is finally reached. In general, we see that there is a slight
decrease of $\alpha_S(M_Z^2)$ from our standard value of 0.1197 to
0.1178.

\subsection{Effect of the choice of the cut on
$Q^2$}\label{sec:effectQsquared}

Repeating the above procedure, but now performing fits with
different choices of the $Q^2$ cut, we obtain the results shown in
Table~\ref{tab:t2}.
\begin{table}[ht]\begin{center}
\begin{tabular}{|c|c|c|c|c|c|c|} \hline
$Q^2_{\rm cut}\,(\GeV^2):$ & 2 & 4 & 7 & 10 & 14 & 20 \\
\hline \# data points & 2097 & 1868 & 1681 & 1537 & 1398 & 1244 \\
$\alpha_S(M_Z^2)$ & 0.1197 & 0.1194 & 0.1185 & 0.1180 &
0.1169 & 0.1174  \\ \hline $\chi^2 (Q^2>2)$ & 2267 &&&&& \\
$\chi^2(Q^2>4)$ & 2046 & 2022 &&&& \\
$\chi^2(Q^2>7)$ & 1844 & 1824 & 1806 &&& \\
$\chi^2(Q^2>10)$ & 1716 & 1694 & 1670 & 1656 && \\
$\chi^2(Q^2>14)$ & 1594 & 1573 & 1553 & 1536 & 1533 & \\
$\chi^2(Q^2>20)$ & 1406 & 1388 & 1370 & 1354 & 1351 & 1348 \\
\hline $\Delta_i^{i+1}$ & \multicolumn{6}{| l |}{$\qquad\ \   0.11
\quad\ \ \,  0.10 \quad\ \ \;  0.10 \quad\ \ \ \, 0.02 \quad\ \ \
0.02
\quad\   $} \\
\hline
\end{tabular} \caption{\label{tab:t2} Each column shows the $\chi^2$ values
obtained from a NLO global analysis
with a different choice of the cut in $Q^2$, together with the
number of data points fitted and the value of $\alpha_S(M_Z^2)$
obtained. The first $\chi^2$ entry in a given column is the total
$\chi^2$, and the subsequent entries show the contributions to
$\chi^2$ from subsets of the data that are fitted.
$\Delta_i^{i+1}$, shown in the final row, is a measure of
stability to changing the choice of $Q^2_{\rm cut}$, as explained
in the text.}
\end{center}
\end{table}
Here the behaviour of the values of $\Delta_i^{i+1}$ is not so
dramatic as for the study of $x_{\rm cut}$. As a consequence it is
more difficult to select the value of $Q^2_{\rm cut}$ for which
the theoretical description becomes safe. However, there is a
general decrease in the value of $\Delta_i^{i+1}$ as the value
$Q^2_{\rm cut}$ is increased. Certainly the choice $Q^2_{\rm
cut}<7~\GeV^2$ appears inappropriate, and $Q^2_{\rm
cut}>14~\GeV^2$ is definitely acceptable. We therefore take the
reasonably conservative choice of $Q^2_{\rm cut}=10~\GeV^2$. This
gradual reduction of $\Delta_i^{i+1}$ is an indication of the
presence of higher-order corrections, whose relative strength
falls off only like $1/\ln Q^2$. If higher-twist corrections were
the dominant effect, then we would expect a more dramatic
reduction of $\Delta_i^{i+1}$ at some low value of $Q^2_{\rm
cut}$.

\subsection{Effect of a cut on the product $xQ^2$}
\label{sec:effectproduct}

The theoretical uncertainties at small $x$ and small $Q^2$ may be strongly correlated.
That is, the main theoretical uncertainty at small $x$ is due to higher-twist effects rather than higher-order
contributions. In order to investigate this possibility, we perform a series of global fits, each with a different
choice of cut on the product $xQ^2$. We begin with $(xQ^2)_{\rm cut} = 0.001~\GeV^2$, which corresponds to a loss
of 42 data points, and proceed in steps of a loss of about a further 50--100 data points until stability is
achieved. This occurs when $(xQ^2)_{\rm cut} = 0.6~\GeV^2$, at which point 589 data points have been removed. The
corresponding value of $\alpha_S(M_Z^2) = 0.1183$, and the gluon distribution, are both very similar to those
obtained with $x_{\rm cut} = 0.005$. Hence we arrive at a similar final result to that of applying an $x_{\rm
cut}$ alone, but with the loss of about twice as much data. Indeed we have removed all HERA data below
$x\sim0.003$, and have only a handful of high $Q^2$ points left for $x<0.005$. Thus to cut on $xQ^2$ appears to be
much more inefficient than to cut on $x$ alone.

\subsection{Combined cut on $x$ and $Q^2$}
\label{sec:effectcombined}

Finally we check the effect of imposing a range of cuts on $x$ having already chosen the conservative $Q^2$ cut of
$Q^2_{\rm cut} = 10~\GeV^2$, and alternatively choosing different cuts in $Q^2$ having already selected the safe
$x_{\rm cut}$ of 0.005. In each case the additional cut leads to further improvements in the quality of the fit,
and complete stability is only achieved when the combined cuts $x_{\rm cut} = 0.005$ and $Q^2_{\rm cut} =
10~\GeV^2$ are imposed. ($\Delta^{i+1}_i = 0.13$ when going from $x_{\rm cut} = 0.0025$ to $0.005$ at $Q^2_{\rm
cut}=10~\GeV^2$, and $\Delta^{i+1}_i = 0.05$ when going from $Q^2_{\rm cut} =7~\GeV^2$ to $10~\GeV^2$ at $x_{\rm
cut}=0.005$). This combined cut yields, in our opinion, parton distributions largely free from theoretical
uncertainties, but only within this restricted kinematic domain. We denote this conservative parton set by
MRST(cons). The corresponding value of $\alpha_S(M_Z^2)$ is 0.1162, and the gluon distribution is similar in shape
to that obtained for the fit with $x_{\rm cut} = 0.005$ (and $Q^2_{\rm cut} = 2~\GeV^2$). To estimate the error of
this MRST(cons) prediction of $\alpha_S$ we use a tolerance of $\Delta\chi^2 = 5$, instead of our previous choice
of $\Delta\chi^2=20$~\cite{MRST2002}, since we now have a more compatible collection of data, without so many
accompanying `tensions' between different data sets. 
However, the reduced amount of fitted data with, in
particular, a more limited range of $Q^2$, yields an intrinsically larger error so that our final uncertainty on
$\alpha_S$ is again approximately $\pm0.002$ (despite the smaller tolerance).

\begin{figure}[htbp]
\epsfig{figure=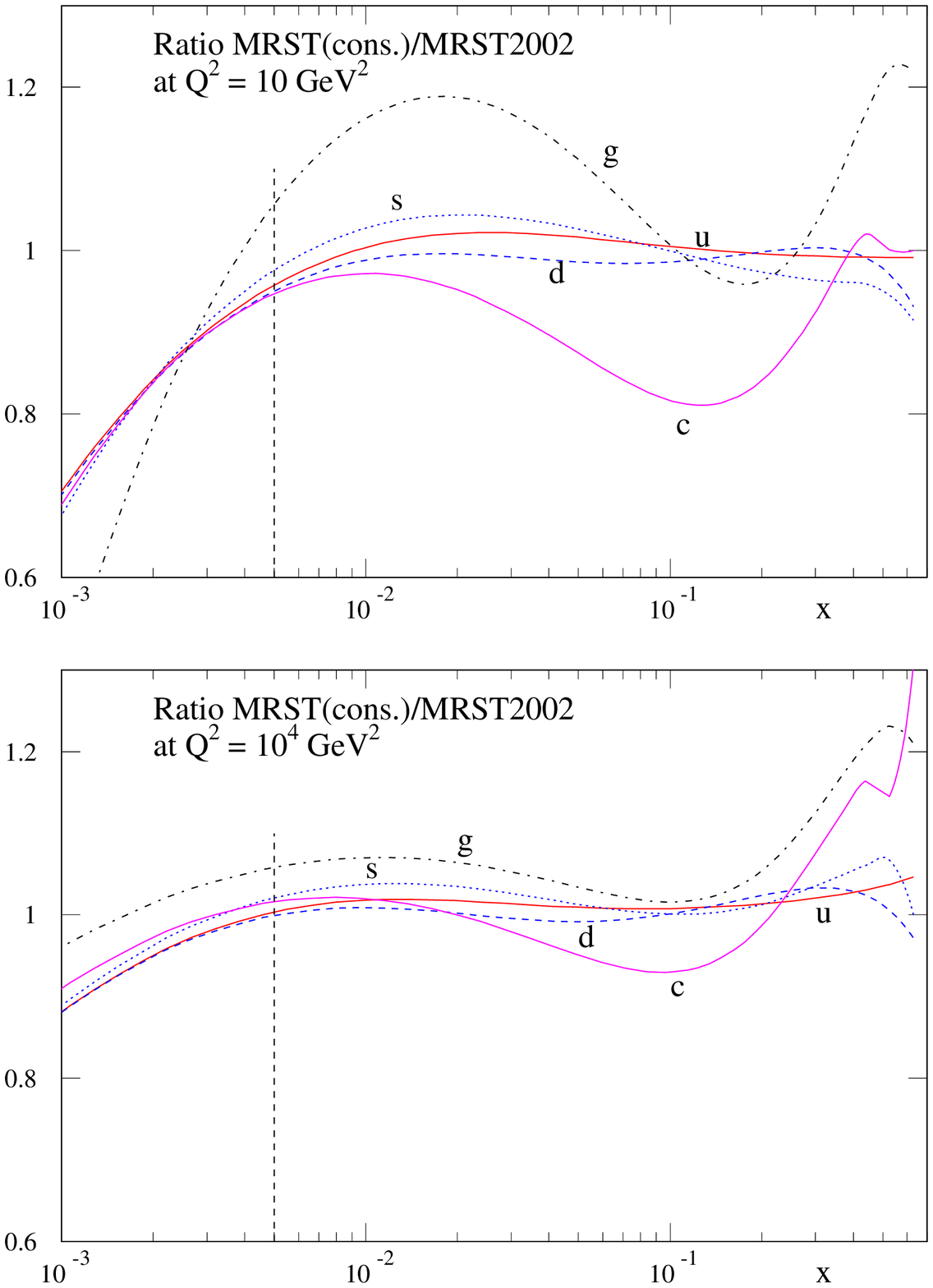,height=20cm}
 \caption{\label{fig:f2} Comparison of MRST(cons) partons with the default NLO set
 MRST(2002). The former partons are only reliable for $x>0.005$
 and $Q^2>10~\GeV^2$.}
\end{figure}

Fig.~\ref{fig:f2} compares the MRST(cons) partons distributions
with those of our default set.
We stress that these partons have a limited range of
applicability. In order to investigate the theoretical uncertainty
outside the kinematic domain, $x>0.005$ and $Q^2>10~\GeV^2$, of
the fitted data, we next study a variety of possible theoretical
corrections to the NLO leading-twist framework.

\section{Uncertainty associated with change from NLO to NNLO fit}
\label{sec:change}

Clearly there will be uncertainties associated with the truncation
of the DGLAP evolution at a given perturbative order. In the past,
these have usually been estimated by seeing the effect of changes
of scale on physical observables in the NLO analysis.
However, this approach is flawed since it provides no information about
higher logs in $1/x$ and $(1-x)$ at higher orders, and now we can
do better. The deep inelastic coefficient functions are known at
NNLO~\cite{CF}. Moreover, valuable partial information has been
obtained about the NNLO splitting functions~\cite{SF,S47}. This
greatly limits the possible behaviour of the splitting functions
down to quite small values of $x$. Indeed, van~Neerven and
Vogt~\cite{VV12} have constructed a range of compact analytic
functions that are all compatible with the available information.

\subsection{Effect of NNLO corrections}

We have performed NNLO global analyses in two previous publications \cite{MRSTNNLO1, MRSTNNLO2}. The dominant
effects of the NNLO corrections are indeed the additional $\ln(1-x)$ terms in the nonsinglet coefficient functions
which influence large $x$, and the behaviour of the coefficient functions and splitting functions at small $x$,
which is heavily determined by the leading $\ln(1/x)$ terms. The additional terms in the nonsinglet coefficient
function lead to an enhancement of the large $x$ structure functions, and hence a small, but significant,
reduction in the valence quark distributions at large $x$. However, because this enhancement is proportional to
$\alpha_S^2(Q^2)$ it falls quickly with $Q^2$ and the evolution of the structure functions increases at large $x$.
This enables a slightly better fit to be obtained, but the natural increase in evolution speed results in a lower
value of $\alpha_S(M_Z^2)$, i.e. it falls from $\sim 0.119$ at NLO to $\sim 0.116$ at NNLO, which is mainly
determined by the high-$x$ structure function evolution.

At small $x$ the speed of evolution of $F_2(x,Q^2)$ is also increased, both by the behaviour of the NNLO splitting
function $P^{(2)}_{qg}(x)$ and the NNLO coefficient function $C^{(2)}_{2g}(x)$. The evolution of the gluon is
decreased, however, due to the leading $\ln(1/x)/x$ term in $P^{(2)}_{gg}(x)$ having a negative coefficient. In our
first analysis we found that the small $x$ gluon at NNLO was far below that at NLO, but this was largely due to
the initial estimate of $P^{(2)}_{qg}(x)$, which was reduced significantly when some additional moments became
available. In our more recent analysis we do indeed find that the NNLO gluon is a little smaller at small $x$ due
to the natural increase in evolution of $F_2(x,Q^2)$. However, the large $x$ gluon needs to be larger at NNLO
because the decrease in the large $x$ quarks discussed above reduces the high-$E_T$ Tevatron jet cross section,
which must be corrected by an increase in the large-$x$ gluon distribution.\footnote{There is not, at present, a
full NNLO calculation of the jet cross section available, but a calculation of leading threshold logarithms
\cite{OWKID} suggests that the NNLO contribution is not large or very $E_T$-dependent.} This redistribution of the
gluon between large and small $x$ is qualitatively in agreement with the momentum sum rule, which is a strong
constraint on the form of the gluon. Indeed, the complete NNLO fit works nicely, and displays a very slight
improvement in quality compared to NLO.

However, it is not clear whether additional large logarithms in $(1-x)$ and $1/x$ beyond NNLO will lead to further
modifications to the partons, and any resulting predictions. We have produced predictions for $W$ and $Z$
production at the Tevatron and the LHC in \cite{MRSTNNLO1,MRSTNNLO2}, and also for the longitudinal structure
function $F_L(x,Q^2)$ (the NNLO coefficient function for this  having been estimated \cite{MRSTNNLO1} in the same
way as the NNLO splitting functions). The former are quite stable, showing changes of $\sim 4\%$ when going from
NLO to NNLO. However, this is larger than the $\sim 2\%$ `experimental' error estimated at NLO \cite{CTEQ6,
MRST2002}, and is dependent on the quark distributions, which are directly obtained by comparison with the
structure functions. In contrast, $F_L(x,Q^2)$ is dependent mainly on the gluon which has no direct constraint at
low $x$, so that when one goes from LO to NLO to NNLO the variation is very large, i.e., $\sim 20\%$ at
$x=0.0001$, even for $Q^2
> 100~\GeV^2$. This variation is driven largely by the $\ln(1/x)$ terms, and it is not clear if any sort of
convergence has been reached at NNLO. The same is potentially true for other gluon dominated quantities.

\subsection{Cuts on Data at NNLO}\label{sec:cutsNNLO}

We might hope that at NNLO the cuts on data required to achieve stability are rather less stringent than at NLO.
However, in the same way that the quality of the global fit only improves very slightly going from NLO to NNLO,
the size of the cuts in $W^2$, $Q^2$ and $x$ does not change much. Nevertheless, there are some advantages to be
gained at NNLO as we discuss in detail below.

Raising $W^2_{\rm cut}$ from $12.5~\GeV^2$ has a similar effect as at NLO. However, this stability at $W^2_{\rm
cut} =12.5~\GeV^2$ or at most $15~\GeV^2$ is subject to the same caveat as at NLO, i.e. a certain incompatibility
of data for $W^2 \sim 15~\GeV^2$. It is also true that the low $W^2$ data is one of the major constraints on
$\alpha_S(M_Z^2)$, which is rather lower at NNLO than NLO, being more consistent between different data sets at
NNLO. Also, if one extrapolates the NNLO fit below $W^2=12.5~\GeV^2$ the departure between data and theory occurs
more slowly than at NLO. Hence, although the `optimal' value of $W^2_{\rm cut}$ does not change at NNLO there are
indications that the theoretical description is improving at low $W^2$. We will return to this point later.

\begin{figure}[htbp]
\epsfig{figure=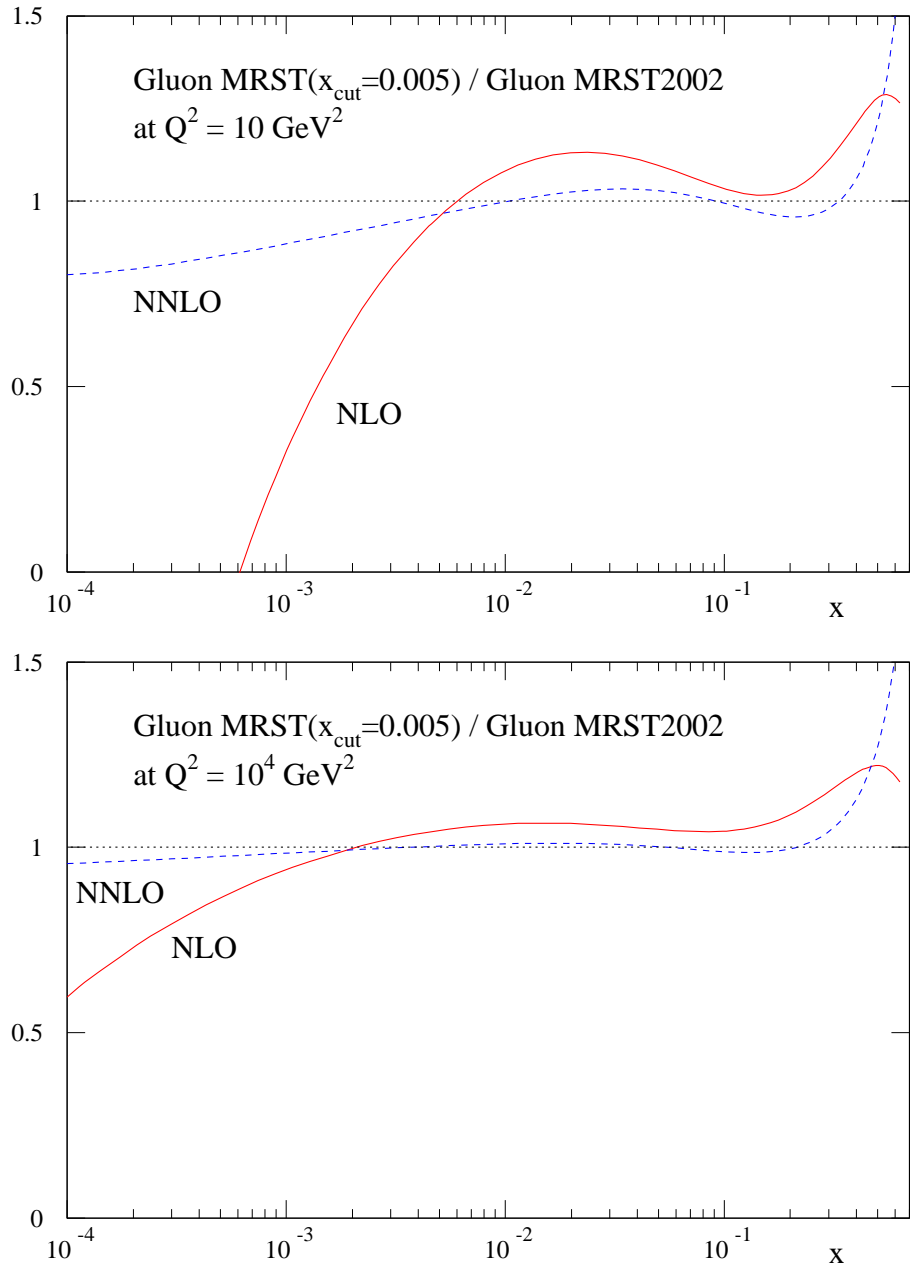,height=20cm}
 \caption{\label{fig:fgluons} Comparison of MRST gluons obtained from
 $x_{\rm cut}=0.005$ with the default gluons
 MRST(2002) at both NLO and NNLO.}
\end{figure}

At NNLO there is a significant improvement each time $x_{\rm cut}$ is raised until $x_{\rm cut}=0.005$, as at NLO.
However, the improvement in going from $x_{\rm cut} = 0.001$  to $x_{\rm cut} = 0.005$ is not as great as at NLO.
The dominant improvement in the quality of the fit is due to a much improved fit to the Tevatron jet data due to a
readjustment of the high $x$ gluon. This was also the case at NLO, but there was a far more significant
improvement in the description of the NMC data and the $x>0.005$ HERA structure function data in that case. As at
NLO, each time the value of $x_{\rm cut}$ is raised there is a reduction in the very low $x$ gluon, and a
corresponding general increase in the high and moderate $x$ gluon. However, the effect is far less pronounced at
NNLO than it was at NLO, as seen in Fig.~\ref{fig:fgluons} which compares the default gluons to the $x_{\rm
cut}=0.005$ versions at both NLO and NNLO. Similarly, while $\alpha_S(M_Z^2)$ decreased as $x_{\rm cut}$ was
lowered at NLO, it is almost completely insensitive to $x_{\rm cut}$ at NNLO. Hence, although the value of $x$
above which we have complete confidence in our partons is the same at NNLO as at NLO, the changes compared to the
default set, and hence the uncertainties involved with the default set, are much reduced at NNLO.

It is a similar story for $Q^2$ cuts. As at NLO there is a continual improvement until $Q^2_{\rm cut} =
10~\GeV^2$, above which stability sets in, and as at NLO this improvement is relatively gradual between $Q^2_{\rm
cut} = 2~\GeV^2$ and $Q^2_{\rm cut} = 10~\GeV^2$, suggesting that higher twist is not most obviously the remaining
correction to theory required. However, as with $x$ cuts, the total improvement in the quality of the fit and the
degree of readjustment of the partons is not quite as large as at NLO.

\begin{figure}[htbp]
\epsfig{figure=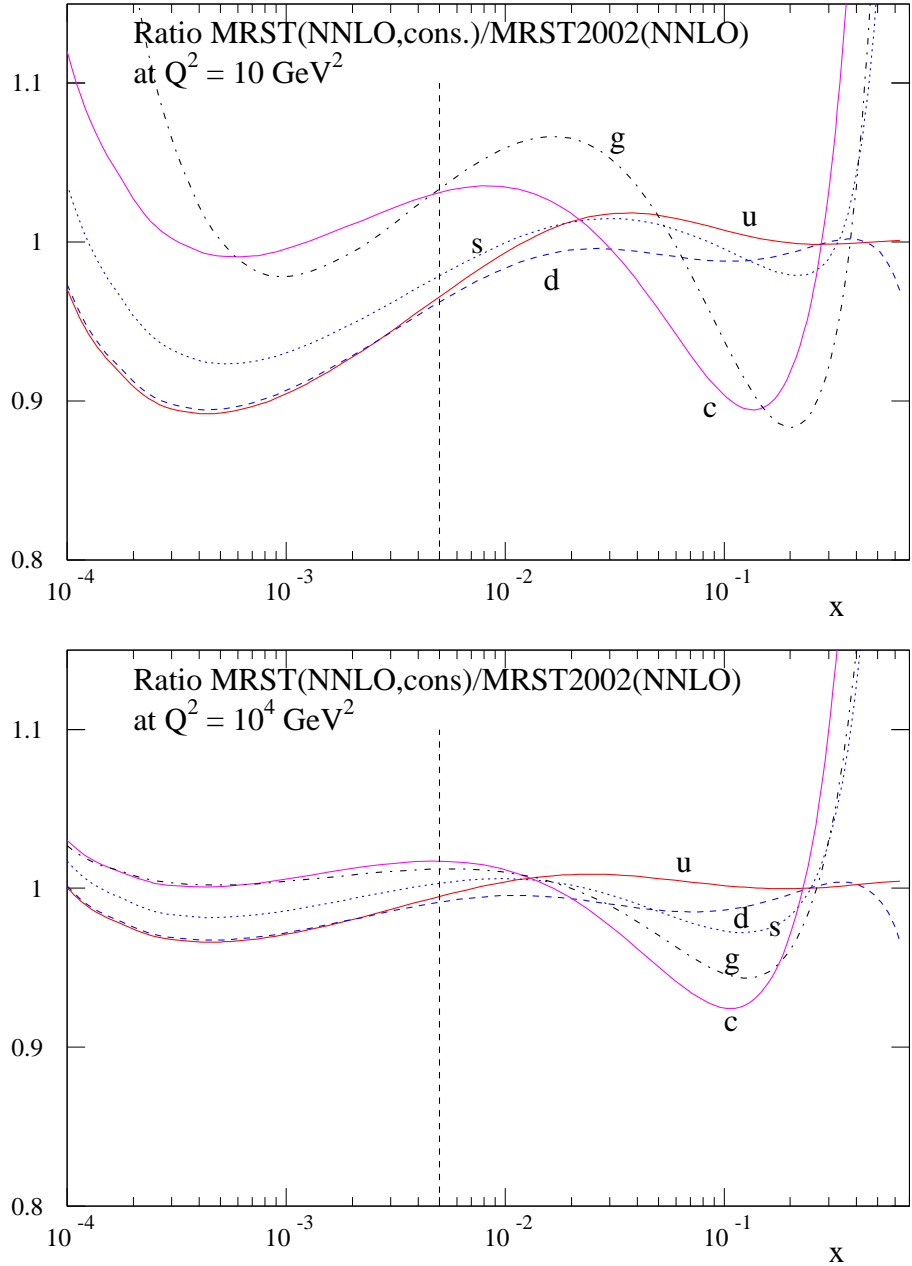,height=20cm}
 \caption{\label{fig:NNLOcon} Comparison of the NNLO MRST(cons) partons with the default set
 MRST(2002) at NNLO. The former partons are only reliable for $x>0.005$
 and $Q^2>7~\GeV^2$.}
\end{figure}

Finally, we investigate the effects of cuts on both $x$ and $Q^2$. With our experience at NLO we would expect
stability to be achieved when both $x_{\rm cut}$ and $Q^2_{\rm cut}$ are near their respective individual values
for stability. However, there is a slight improvement at NNLO as compared to NLO. In this case $\Delta^{i+1}_i$ is
negligible when going from $Q^2_{\rm cut} =7~\GeV^2$ to $10~\GeV^2$ at $x_{\rm cut}=0.005$. However,
$\Delta^{i+1}_i = 0.12$ when going from $x_{\rm cut} = 0.0025$ to $0.005$ at $Q^2_{\rm cut}=7~\GeV^2$, and
similarly if $Q^2_{\rm cut}=10~\GeV^2$. Hence, our conservative set of NNLO partons is obtained with the slightly less
severe $Q^2_{\rm cut}=7~\GeV^2$ and $x_{\rm cut} = 0.005$. We also note that the modification of the gluon
distribution for the MRST(cons) set is far smaller at NNLO than at NLO, being similar to the $x_{\rm cut}=0.005$
alone set as shown in Fig.~\ref{fig:fgluons}, and the value of $\alpha_S(M_Z^2)$ reduces only to $0.1153\pm 0.002$, a
much smaller reduction than at NLO. As at NLO, the error corresponds to the reduced tolerance $\Delta\chi^2 =5$.
The detailed comparison of the `conservative' set with
the default set of partons is shown in Fig.~\ref{fig:NNLOcon},
and one sees that the deviation is rather smaller than at NLO, particularly at small $x$.

Hence, the investigation of the full range of cuts on the data that are fitted results in a set of
`conservative' NNLO partons which are only completely reliable over a slightly extended range compared to NLO.
However, the change in these partons compared to the default set within this range, and more especially outside
the range, is much smaller than at NLO (and the same is true for the change in $\alpha_S(M_Z^2)$). This implies
that predictions made using NNLO partons are considerably more reliable than those using NLO partons, even before
we consider the extra precision obtained simply by using the expressions for cross sections at a higher order. We
will discuss this in more detail later.

\section{Specific theoretical uncertainties} \label{sec:specific}

The results of Sections~\ref{sec:selection} and \ref{sec:change}
imply that there may be significant theoretical corrections beyond
NNLO in the standard DGLAP perturbative expansion, which become
important at either low $x$, low $Q^2$ or low $W^2$ or some
combination of these. Indeed there exist several types of
theoretical correction which may be expected to have such effects.
These include
\begin{itemize}
\item[(i)] Higher powers of $\ln 1/x$
at higher orders in $\a$, which become important at low $x$.
\item[(ii)] Increasing powers of $\ln(1-x)$ at higher orders in
$\a$, which are generally well understood, but which are intrinsically linked to higher-twist corrections. These
are important at high $x$ and hence low $W$, see~(\ref{eq:Wsquared}).
\item[(iii)] Absorptive
corrections which arise from parton recombination and which are
higher twist in nature. These should become important at low $x$
and $Q^2$.
\item[(iv)] Residual higher-twist contributions, which
will be important at low $Q^2$. In practice these are often combined with the specific higher-twist contributions,
already mentioned, in a purely phenomenological parameterization.
\end{itemize}

We have performed separate global analyses to investigate the
effect of each of these in turn. The results are described below.

\subsection{Contribution of higher order $\ln 1/x$
terms}\label{sec:1/x}

It has long been known that the splitting and coefficient functions typically contain one additional power of $\ln 1/x$ for
each additional power of $\a$. Many of these terms are known explicitly~\cite{BFKL,CH,BFKLNLL}. However, a full
$\ln 1/x$ resummation seems to involve many complications; for example, treatment of the running coupling,
kinematic constraints, collinear resummations, etc. Various procedures exist for incorporating 
a full $\ln 1/x$ resummation but there is, as yet, no agreed prescription.

Hence, in the present study, we take a general approach. We
include higher order corrections to the NLO splitting functions,
with the correct maximum power of $\ln 1/x$, but we let the
coefficient be determined by the global fit to the data. We begin
by adding one additional term to $P_{gg}$ and to $P_{qg}$. For
this investigation we choose to include phenomenological
$\a^4\ln^3 1/x$-type terms of the form
\be P_{gg} \ra P_{gg}^{\rm NLO} + A\ol\a^4
\left(\frac{\ln^31/x}{3!} - \frac{\ln^21/x}{2!}\right)
\label{eq:addtoPgg} \ee
\be P_{qg} \ra P_{qg}^{\rm NLO} + B\a \frac{n_f}{3\pi}\ol\a^4
\left(\frac{\ln^31/x}{3!} - \frac{\ln^21/x}{2!}\right),
\label{eq:addtoPqg} \ee
where $\ol\a = 3\a/\pi$ and $n_f$ is the number of active quark
flavours. Both of the additional terms have been constructed so
that momentum is still conserved in the evolution. We also add the
same terms multiplied by $C_F/C_A = 4/9$ to $P_{gq}$ and $P_{qq}$
respectively. This factor of $C_F/C_A$ is typical for the results
in $\ln 1/x$ resummation~\cite{CH}.

The best  NLO global fit, modified as in (\ref{eq:addtoPgg}) and (\ref{eq:addtoPqg}), gives an improvement in
$\chi^2$ of 21 compared to MRST2002~\cite{MRST2002} and corresponds to $A=3.86$ and $B=5.12$. These are {\em
effective} coefficients and should not be directly compared with the known coefficients from $\ln 1/x$
resummations, since they represent the effect of the two towers of $\ln^n 1/x$ terms. However, the values of $A$
and $B$ are of the magnitude expected from the partial information which exists. Hence the fit seems to benefit
from such terms, which increase the speed of evolution at small $x$. The best-fit value of 
$\a(M_Z^2)$ decreases, but only very
slightly. The input gluon turns out to be similar to that obtained in the fits when data below $x=0.005$ are
removed; that is, it is larger than the default gluon for large and moderate $x$, but even more negative at very
small $x$. We can only assume that large positive $\ln 1/x$ terms in the coefficient functions will maintain the
positivity of observables sensitive to the gluon, such as $F_L$~\cite{RST}. Nevertheless, the increased gluon
evolution at small $x$ does result in a positive gluon more quickly as $Q^2$ increases.

We also investigated the effect of a more flexible parameterization of the $\ln 1/x$ resummation by introducing an
additional term, both in (\ref{eq:addtoPgg}) and in (\ref{eq:addtoPqg}), of the type
\be C \ol \a^5\left(\frac{\ln^4 1/x}{4!} - \frac{\ln^3
1/x}{3!}\right) \label{eq:Cterm} \ee
\be D \a \frac{n_f}{3\pi} \ol\a^3 \left(\frac{\ln^2 1/x}{2!} -
\ln 1/x\right) \label{eq:Dterm} \ee
respectively. Again the form of the terms has been constructed so
that momentum is conserved in the evolution. With the two
additional parameters, we have an effective parameterization of both the
N$^3$LO and N$^4$LO corrections at small $x$.
The best fit now has a $\chi^2$ that is 36
lower than MRST2002 with
\be A=-0.27, \quad B=2.79, \quad C=4.08, \quad D=1.09.
\label{eq:parametervalues} \ee
The gluon distribution is very similar to the case with just two extra parameters, as is $\a(M_Z^2)$. However,
neither the small $x$ low $Q^2$ gluon distribution or the parameters $A$, $B$, $C$ and $D$ are determined very
precisely, since there is some trade-off between them. For example, since the term in~(\ref{eq:Cterm}) falls off
very quickly with $Q^2$ (since it behaves as $\alpha_S^5(Q^2)$), a more negative input gluon is largely compensated
by a larger value of $C$, resulting in a largely unchanged gluon when one moves far away from the input scale
$Q_0^2=1~\GeV^2$. However, the generally positive input parameters and more negative input gluon is an unambiguous
result. The fact that $\a(M_Z^2)$ is not altered much by the inclusion of terms which are important at small $x$
is not surprising since it is always the evolution of the large $x$ structure functions that has the dominant
influence on the extracted value of $\a(M_Z^2)$, and this is not affected by this type of modification. The
studies with up to four additional parameters were deemed sufficient to illustrate the possible effect of $\ln
1/x$ resummation on the NLO analysis.

\begin{figure}[htbp]
\epsfig{figure=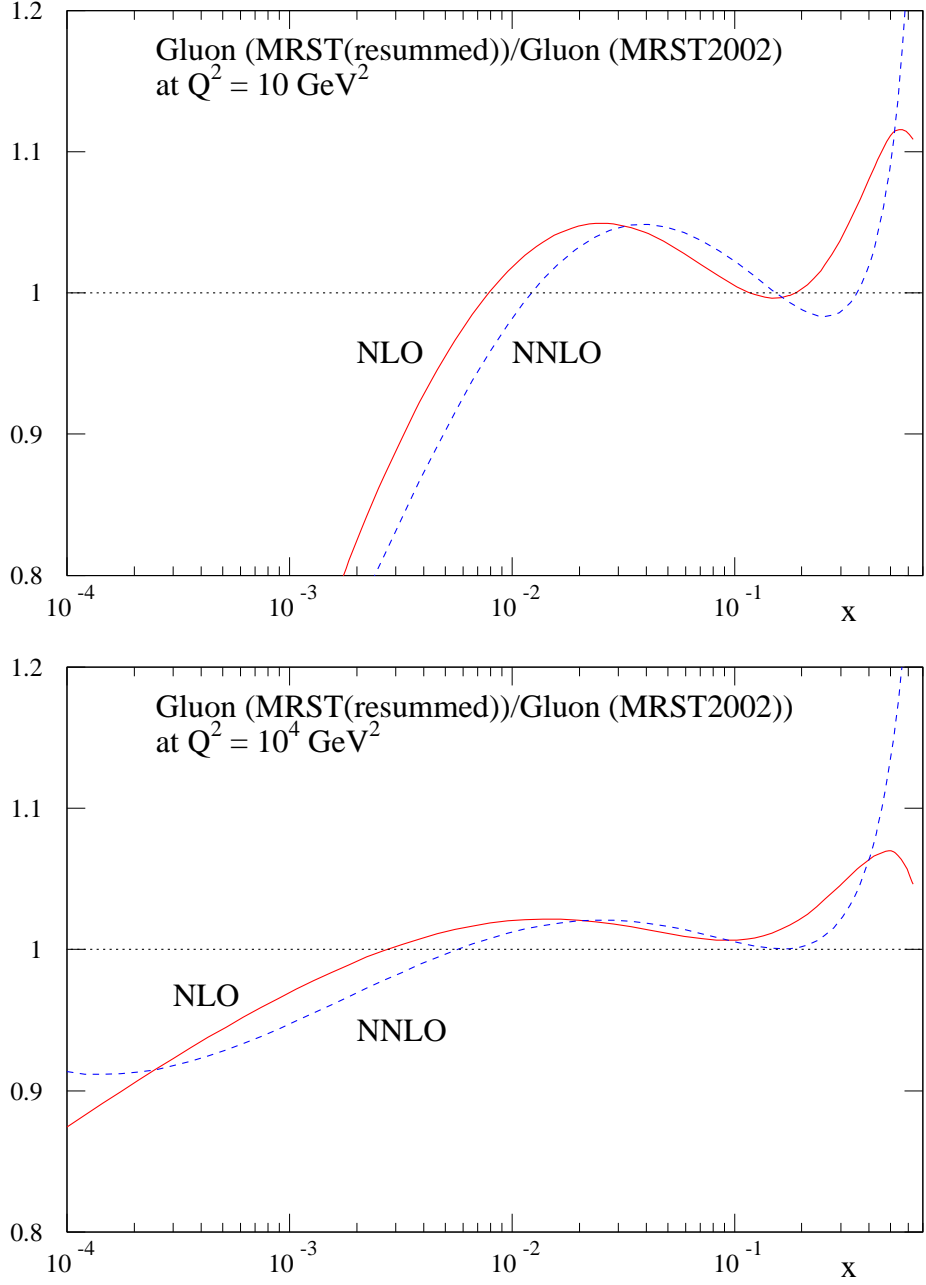,height=8in}
 \caption{\label{fig:xgluons} Comparison of MRST gluons obtained from
 the fits with additional resummation corrections in $\ln(1/x)$
 with the default gluons
 MRST(2002) at both NLO and NNLO.}
 \end{figure}

However, we already know that the approximate NNLO splitting functions have a significant effect on the small $x$
evolution. These contain one extra power of $\ln 1/x$ in comparison to the NLO splitting functions. Hence we
repeated the above analysis at NNLO with the same four parameters, $A,\dots, D$. The results were largely similar.
The $\chi^2$ improved by 39 from the standard MRST(NNLO) fit\footnote{In practice we compared to the NNLO set of partons
corresponding to the MRST2002 analysis~\cite{MRST2002}.}~\cite{MRST2002}, although the parameters take the values
\be A=-0.35, \quad B=-2.00, \quad C=5.49, \quad D=2.81,
\label{eq:parametervalues2} \ee
very different from those at NLO, see (\ref{eq:parametervalues}). In addition they have a markedly reduced positive effect
for the quarks and a very slightly increased positive effect for the gluon, 
showing that the NNLO contributions (positive for 
quark evolution at small $x$ but negative for gluon evolution) have themselves been important at small
$x$.\footnote{The effect is usually presented at quite high $Q^2$, e.g. $30~\GeV^2$, and with partons which are
steep at small $x$ \cite{VV12}, and is claimed to be quite moderate. At $Q^2\sim 5~\GeV^2$, with partons which are
flattish at small $x$, the contribution from the NNLO splitting functions is proportionally much larger, and
certainly very significant.} As at NLO, the gluon at high and moderate $x$ is increased as compared to the
unmodified NNLO fit, while at small $x$ it is more negative. The gluons in these $\ln(1/x)$-modified fits are 
shown in Fig.~\ref{fig:xgluons}. At low $Q^2$ they are indeed both much reduced compared to the default gluons 
at small $x$. However, the additional terms in the gluon splitting functions and the additional 
gluons at moderate $x$  
drive the small-$x$ evolution more quickly than the default and at high $Q^2$ the  $\ln(1/x)$-modified gluons
are only a little lower than the default gluons at small $x$. 

Hence, the distinct quality in the improvement in the global fit provides strong evidence that large $\ln 1/x$
contributions beyond even NNLO may be important in the description
of the data. However, these corrections are empirically different in the NLO case than in NNLO, because the latter
has its own large terms for $x \to 0$.

\subsection{Contribution of higher order $\ln (1-x)$ terms} \label{sec:1-x}

If we expand the quark coefficient function in powers of
$\alpha_S$, i.e.
\be
C_{i,q}(x,Q^2) = \delta(1-x) + \sum_{m=1} \alpha^m_S(Q^2)c^{(m)}_{i,q}(x),
\ee
then the coefficient functions $c^{(m)}(x)$ and their moments $\tilde
c^{(m)}(N)$ contain large logarithms as $x \to 1$ ($N \to \infty$) of
the form
\be
\biggl[\frac{\ln^{k-1}(1-x)}{1-x}\biggr]_+, \qquad \qquad \frac{(-1)^k}{k}
(\ln^k N)
\ee
where $k=1,\dots,2m$. Combining results on soft gluon resummation \cite{softglu} and finite order results
\cite{CF}, Vogt has been able to provide explicit expressions for the first four coefficients \cite{VOGT} in an
expansion of the form
\be \tilde  C_{i,q}(N,Q^2) = 1 + \sum_{m=1}^{\infty}\alpha_S^m(Q^2)( c_{m1}\ln^{2m}N + c_{m2}\ln^{2m-1}N
+c_{m3}\ln^{2m-2}N +c_{m4}\ln^{2m-3}N +\cdots). \label{highx} \ee Hence, we have detailed knowledge of the leading
terms in the coefficient functions for the large $x$ or $N$ limit at all orders, and it is argued in \cite{VOGT}
that a more efficient convergence is achieved if the leading terms are arranged as powers of  $\ln N$
rather than $\ln(1-x)$. In principle all of the terms in (\ref{highx}) should be included. However, the
investigation in \cite{VOGT} suggests that, unless one is at very high $x$, or equivalently very low $W^2$, going
to a finite order is sufficient. Indeed, the suggestion is that for $m=3$ the coefficient function is only
important above $x \sim 0.7$, and those for $m>3$ are only important for $x> 0.8$.

In order to investigate this we performed a fit using the NNLO splitting functions and coefficient functions, and
including also the ${\cal O}(\alpha_S^3)$ contribution to the coefficient function in (\ref{highx}).\footnote{In \cite{VOGT}
it is demonstrated that the four terms in (\ref{highx}) are a very good approximation to the full NNNLO
coefficient function, which can be estimated in detail using the same sort of techniques as in \cite{VV12}.} When
using our $W^2$ cut of $12.5~\GeV^2$, we find that the effect of the ${\cal O}(\alpha_S^3)$ coefficient function
at very high $x$ is almost negligible, since the $W^2$ cut ensures we are at quite high $Q^2$ for very large $x$,
and $\alpha_S^3(Q^2)$ is fairly small. In fact the most significant effect of the ${\cal O}(\alpha_S^3)$
coefficient function is at higher $x$. Since the coefficient function has a vanishing first moment, its positive
effect at very high $x$ must be countered by a negative contribution at lower $x$. In practice it increases the
structure function for $x>0.55$, but decreases it for $x<0.55$. This decrease is not large, but it affects much
more data. In practice the best fit, when including the approximate NNNLO coefficient function, is very slightly
worse than the usual NNLO fit, although the partons and the value of $\alpha_S(M_Z^2)$ are hardly changed at all.

From \cite{VOGT} it is clear that the contributions in (\ref{highx}) beyond ${\cal O}(\alpha_S^3)$ are only
important at even higher $x$. Hence, we conclude that if we use $W^2_{\rm cut} = 12.5~\GeV^2$, there is no
advantage to be gained in including terms beyond the NNLO coefficient function. As we go lower in $W^2$, however,
the NNNLO coefficient function does start to have a non-negligible effect. We will discuss this in more detail in
our analysis of higher-twist corrections.

Indeed, with reference to higher twist, we should note that there are a number of ambiguities when performing
large $x$ resummations, not just whether to resum large logarithms in $N$ or $(1-x)$. We note that the series
expansion in (\ref{highx}) is convergent. However, one could alternatively calculate the first two towers of terms
in the expansion
\be \frac { d \ln \tilde C_{i,q}(N,Q^2)}{d \ln Q^2} = \sum_{m=1}^{\infty}\alpha_S^m(Q^2)( c_{m1}\ln^{m}N +
c_{m2}\ln^{m-1}N +\cdots), \label{highxderiv} \ee which would give the dominant large-$N$ contribution to an
effective anomalous dimension for structure function evolution. In this case the series would have finite radius
of convergence. This badly-defined series shows that there are higher twist corrections present, and the ambiguity
in the series is taken as an estimate of the size of the higher-twist corrections in renormalon models. The divergence
in (\ref{highxderiv}) is reflected in the terms in (\ref{highx}) beyond $c_{m4}$ becoming extremely large.
Strictly speaking one cannot simply perform a large $\ln(1-x)$ expansion without encountering this interplay with
higher-twist corrections, and beyond about NNNLO it is difficult to disentangle the two.

\subsection{Absorptive effects} \label{sec:absorptive}

To investigate the effects of absorption\footnote{We thank
M.G.~Ryskin for discussions.} we include bilinear terms in
evolution equations as follows
\be \frac{\partial (xg(x,Q^2))}{\partial\ln Q^2} = \dots -
3\frac{\a^2(Q^2)}{R^2Q^2} \int_x^1 \, \frac{dx'}{x'}\,
[x'g(x',Q^2)]^2 \label{eq:bilineareveq1} \ee
\be \frac{\partial(xq(x,Q^2))}{\partial\ln Q^2} = \dots -
\frac{1}{10}\frac{\a^2(Q^2)}{R^2Q^2} [xg(x,Q^2)]^2
\label{eq:bilineareveq2} \ee
These terms take into account the dominant small $x$ contributions at lowest order in $\a$, as calculated by
Mueller and Qiu~\cite{MQ}. We have approximated the two-gluon correlation function as
\be g^{(2)}(x,Q^2)) =
\frac{2}{3\pi^2R^2}[g(x,Q^2)]^2 \label{eq:twogluon} \ee
as estimated in \cite{MQ}. We consider two choices
of $R^2$, namely $R^2=15$ and $5~\GeV^{-2}$. The former represents
the na\"{\i}ve assumption that $R$ is of the order of the proton
radius, whereas the latter represents much stronger absorption
motivated by `hotspot' studies which allow a large contribution from the
diagrams responsible for saturation where both gluon ladders couple to the
same parton~\cite{LR}.
 Of course, inclusion of the
absorptive terms leads to a violation of momentum conservation. In
principle a more complete theoretical treatment would correct for
this, but in this study we account for the effect by starting the
evolution at low scales with the partons carrying slightly more
than 100\% of the proton's momentum. The precise amount is chosen
so that at high $Q^2$, when the absorptive corrections have
completely died away, the correct value of 100\% is obtained. In
practice for $R^2=15$ and $5~\GeV^{-2}$ we input a total momentum
of 100.7\% and 103\% respectively at $Q_0^2=1~\GeV^2$.

The effect of absorptive corrections in global parton analyses was
investigated many years ago~\cite{KMRS}. The corrections were
found to have a sizable, observable effect in the small $x$
region accessible to HERA for a parton set B$_-$ in which the
gluon was assumed to have a small $x$ behaviour of the form $xg
\sim x^{-\frac{1}{2}}$ at the then input scale of $Q_0^2=4\
\GeV^2$. However, with the advent of the HERA data, the size of
the small $x$ gluon is now known to be considerably smaller than
that of the B$_-$ set, and hence we may anticipate the shadowing
effects will be much smaller.

Here we make two alternative studies. First, we repeat the NLO global analysis, with the modifications shown in
(\ref{eq:bilineareveq1}) and (\ref{eq:bilineareveq2}), starting from MRST2002 partons~\cite{MRST2002}. For the
$R^2=15~\GeV^{-2}$ choice there is almost no change in the fit, the best fit having an essentially identical $\chi^2$.
There is no significant change in the partons and $\a(M_Z^2)$ increases by less than 0.001. For $R^2=5~\GeV^{-2}$
there is a slight deterioration in the best fit of $\Delta\chi^2\simeq20$. This is accompanied by an increase in
$\a(M_Z^2)$ to 0.1225 and a slight increase in the small $x$ gluon. This shows that the slowing of the evolution
by the absorptive corrections is not consistent with the data, and the increase in $\a$ is necessary to compensate
partially for this.

However, starting from the MRST2002 partons, which have a negative
gluon for low $x$ and $Q^2$, may be regarded to be inconsistent
with an absorptive approach, which really assumes a positive
input. Hence, as an alternative investigation, we force the input
gluon to be positive-definite and, indeed, slowly increasing with
decreasing $x$. Since we know that this causes conflict with the
low $Q^2$ data we also raised the $Q^2$ cut to $Q_{\rm
cut}^2=5~\GeV^2$. However, after performing the fits we
investigated the extrapolation to lower $Q^2$.

\begin{figure}[htbp]
\epsfig{figure=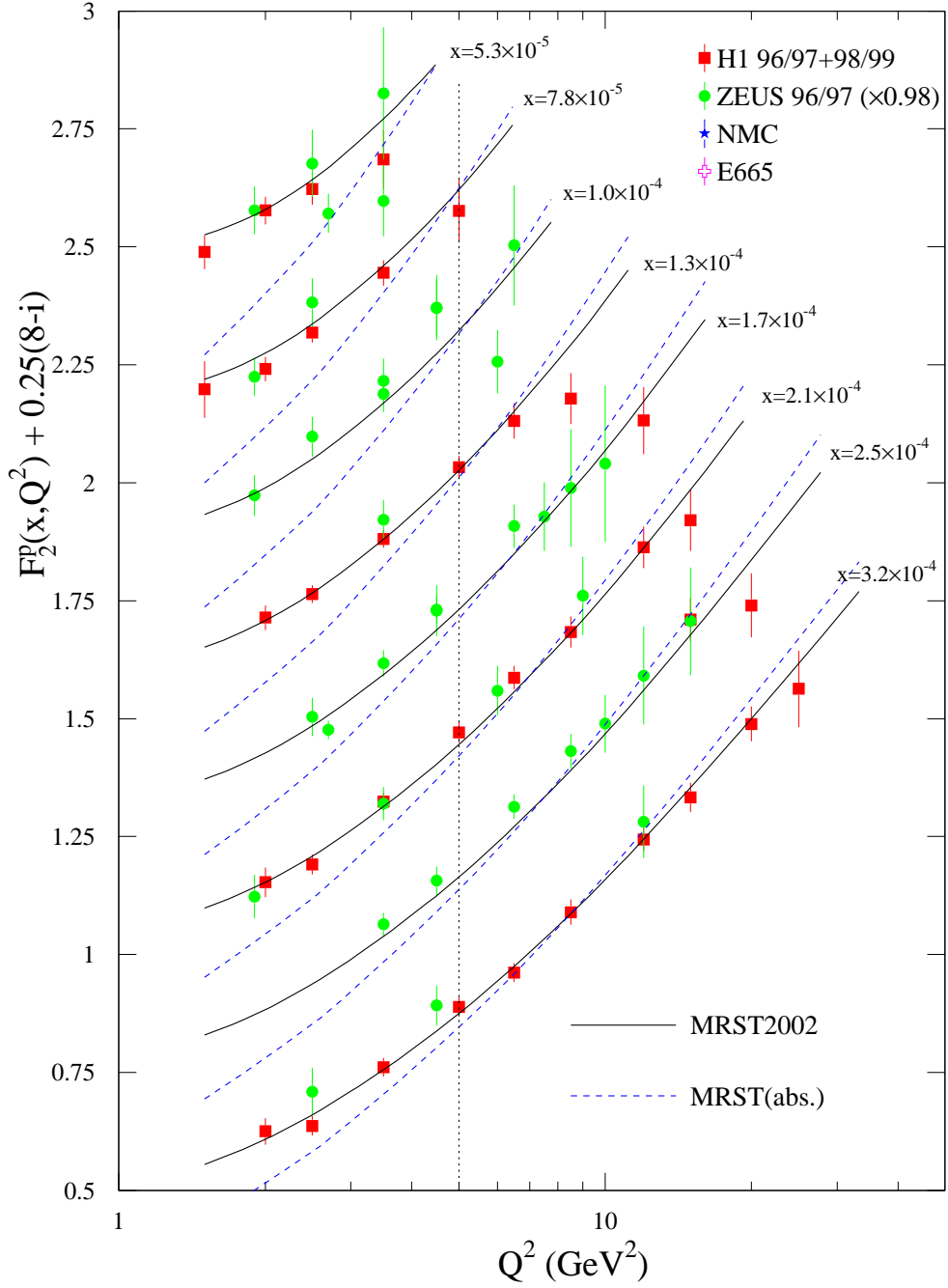,height=8in}
 \caption{\label{lowq} The comparison with data of the fit with a
flattish input gluon and saturation corrections with the curves
extrapolated below the region of validity of the fit.}
 \end{figure}

In detail, we removed the `negative' input term
$[-A_-(1-x)^{\eta_-}x^{-\delta_-}]$ from the gluon, and set the
(conventional) small $x$ power $\delta_g=-0.1$, hence ensuring a
positive definite starting gluon\footnote{It was also necessary to
fix $\varepsilon_g$ to some value, which in practice was chosen to
be 0.74, in order to prevent the interplay between parameters
resulting in an effectively valence-like input gluon
distribution.}. For $R^2=15~\GeV^{-2}$ the best fit, for data
above $Q^2 = 5~\GeV^2$, has $xg(x,Q_0^2)\sim 0.87x^{-0.1}$
at small $x$ and is over 200 worse in $\chi^2$ than
MRST2002, most of this deterioration coming from the HERA data.
The evolution at low $x$ is far too rapid, even though $\a(M_Z^2)$
reduces to 0.117. However, there is also some worsening to the fit
to high $x$ data due to the gluon at high $x$ being smaller than
before. If we extrapolate the fit below $Q^2=5~\GeV^2$ then the
description of the HERA, and even NMC, data becomes very poor indeed, see
Fig.~\ref{lowq}.

For $R^2=5~\GeV^{-2}$ the increased absorptive corrections moderate the problems at lowest $x$ and
$xg(x,Q_0^2)\sim 0.93x^{-0.1}$, i.e., a little larger, and the global $\chi^2$ is 80 worse than for MRST2002 with
$\a(M_Z^2) = 0.118$. However, the extrapolation to low $Q^2$ is still nearly as bad as in the previous case.

Hence we conclude that if we demand a positive-definite input
gluon, which is even slowly increasing as $x$ decreases, then
absorptive corrections are not sufficient to compensate the faster
evolution when compared to a negative input (MRST2002)
gluon\footnote{This result is in conflict with observations of
Ref.~\cite{EHKQS} in which only LO partons are considered.}, and
also the necessary reduction of the larger $x$ gluon has a
detrimental effect on the fit.

It was hoped that the introduction of absorptive corrections would lessen, or perhaps remove, the need for a
negative gluon at very low $x$ and low $Q^2$. However we see from the above studies that the global fits do not
appear to favour the introduction of absorption corrections. This result is surprising~\cite{RY}. We know that at
low $x$ about 10\% of $F_2$ arises from diffractive events. However we cannot simply subtract $F_2^D$ from the
inclusive $F_2$, since the diffractive events which originate from low scales are already accounted for in the
parameterization of the starting distributions at $Q_0$. The diffractive contribution, $\Delta F_2^D$, arising
from higher scales is however related to the absorptive correction $\Delta F_2$ to the inclusive $F_2$. The
relation is given by the AGK cutting rules~\cite{AGK}. When $\Delta F_2 \ll F_2$, the relation takes the form
\be \Delta F_2 \simeq -\Delta F_2^D. \ee
So some absorptive correction is expected to be present. To quantify the amount will require an enlarged global
analysis incorporating the diffractive structure function data.

\subsection{Higher-twist effects} \label{sec:highertwisteffects}

In previous papers (\cite{MRSTtwist,MRSTNNLO1}) we have examined the effect of including in global fits a simple
phenomenological parameterization of the higher-twist contribution in the form
\be F^{\rm HT}_i(x,Q^2) =F^{\rm LT}_i(x,Q^2)\left( 1+\frac{D_i(x)}{Q^2}\right), \label{HT} \ee
where in practice $D_i(x)$ is taken to be a constant, independent of $Q^2$, in each of a number of different bins in $x$.
In the fits which include such a parameterization, we have lowered the $Q^2$ cut to $1.5~\GeV^2$ and the
$W^2$ cut to $4~\GeV^2$. Here we repeat the procedure for our NLO and NNLO fits with the most up-to-date data, and
also include a fit which has the approximate NNNLO coefficient function, as discussed in Section~4.2. This only
alters the NNLO results at high $x$. We give results for a LO fit, although we have already seen in
\cite{MRSTNNLO1} that such a fit simply fails in many regions of parameter space, and the invoked higher-twist
corrections are simply mimicking, as best as they can, corrections which really reproduce the NLO (and possibly
higher) leading-twist contributions. In particular, we see that the large negative higher-twist corrections at
small $x$ decrease significantly at NLO and even tend to disappear altogether at higher orders. The results are
summarised in Table~\ref{tab:t3}.

\begin{table}[h]
\begin{center}
\begin{tabular}{|c|r@{.}l|r@{.}l|r@{.}l|r@{.}l |} \hline
& \multicolumn{2}{c|}{} & \multicolumn{2}{c|}{} & \multicolumn{2}{c|}{} & \multicolumn{2}{c|}{}\\
$x$ & \multicolumn{2}{c|}{LO} & \multicolumn{2}{c|}{NLO} & \multicolumn{2}{c|}{NNLO} & \multicolumn{2}{c|}{NNNLO}\\
\hline
     0--0.0005 & $-0$&$07 $ & $-0$&$02 $ & $-0$&$02 $ & ~$-0$&$03$ \\     
0.0005--0.005  & $-0$&$03 $ & $-0$&$01 $ & $ 0$&$03 $ & $ 0$&$03$ \\
 0.005--0.01   & $-0$&$13 $ & $-0$&$09 $ & $-0$&$04 $ & $-0$&$03$ \\
  0.01--0.06   & $-0$&$09 $ & $-0$&$08 $ & $-0$&$04 $ & $-0$&$03$ \\
  0.06--0.1    & $-0$&$02 $ & $ 0$&$02 $ & $ 0$&$03 $ & $ 0$&$04$ \\
   0.1--0.2    & $-0$&$07 $ & $-0$&$03 $ & $-0$&$00 $ & $ 0$&$01$ \\
   0.2--0.3    & $-0$&$11 $ & $-0$&$09 $ & $-0$&$04 $ & $ 0$&$00$ \\
   0.3--0.4    & $-0$&$06 $ & $-0$&$13 $ & $-0$&$06 $ & $-0$&$01$ \\
   0.4--0.5    & $ 0$&$22 $ & $ 0$&$01 $ & $ 0$&$07 $ & $ 0$&$11$ \\
   0.5--0.6    & $ 0$&$85 $ & $ 0$&$40 $ & $ 0$&$41 $ & $ 0$&$39$ \\
   0.6--0.7    & $ 2$&$6  $ & $ 1$&$7  $ & $ 1$&$6  $ & $ 1$&$4$  \\
   0.7--0.8    & $ 7$&$3  $ & $ 5$&$5  $ & $ 5$&$1  $ & $ 4$&$4$  \\
   0.8--0.9    & $20$&$2  $ & $ 16$&$7 $ & $ 16$&$1 $ & $13$&$4$  \\ \hline
\end{tabular}
\caption{\label{tab:t3}The values of the higher-twist coefficients $D_i$ of (\ref{HT}), in the chosen bins of $x$,
extracted from the LO, NLO, NNLO and NNNLO (NNLO with the approximate NNNLO non-singlet quark coefficient
function) global fits.}
\end{center}
\end{table}

There are a number of conclusions which can be drawn from the Table. First, we can see that there is no clear
evidence for any significant higher-twist contributions for $x<0.005$. Even though it may be argued that the form of the higher-twist
corrections at small $x$ is rather more complicated than the simple parameterization of (\ref{HT}) (e.g.
\cite{Bartels}), they would have to be of roughly the same qualitative form, and the lack of any indication of
them appears compelling.

At NLO there is a strong indication of a negative higher twist contribution for $x\sim 0.005\!-\!0.06$. This is
required in order to make $dF_2(x,Q^2)/d\ln Q^2$ large enough for the NMC data in this region. However, the sizes
of the $D_i$'s in this region decrease significantly when going to NNLO, because both the coefficient functions
and splitting functions at NNLO lead to increased evolution in this range, and the evidence for higher twist at
NLO seems to be really an indication of a lack of important leading-twist, higher-order corrections.

The main higher-twist corrections appear, as expected, at high $x$. For $x\sim0.1\!-\!0.4$ there is a slight
indication of a negative higher-twist correction at NLO, but this diminishes at NNLO and effectively disappears at
NNNLO. Hence this is presumably just an indication that leading-twist perturbative corrections are
important. A transition is apparent for $x\sim 0.4\!-\!0.5$,  and for $x>0.5$ there is a definite positive
higher-twist contribution. However, this contribution has a tendency to decrease from one order to the next.
Indeed at NNNLO the required higher-twist contribution at very high $x$ is similar to that achieved simply from
target-mass corrections \cite{TargetMass}. Moreover, in this very high $x$ domain, the correction to the
higher-twist coefficient when going from NNLO to NNNLO is as large, if not larger, as when going from NLO to NNLO
(the NLO to NNLO correction is smaller than the LO to NLO). It is easily verified that this is indeed because the
NNNLO correction to the structure function is as large at NNNLO as at NNLO for this range of $x$ and $Q^2$. This
shows that the divergent high-$x$ perturbative series discussed in Section~4.2 reaches its minimum at about NNNLO,
and this represents the essential ambiguity in this series. Hence the renormalon contribution to the higher twist
\cite{renormalon}, which comes from this ambiguity of the perturbative series, is roughly of the same size as the
NNNLO contribution (at least in the region of parameter space we are probing). Indeed, the prediction for the
higher-twist contribution as a function of $x$ in \cite{renormalon} is very similar to that obtained from the
approximate NNNLO contribution. This implies that at very high $x$ it is pointless to go beyond NNLO (or
certainly NNNLO) in the perturbative series, since at this order the perturbative series and higher-twist
corrections become indistinguishable. It is important to realize, however, that this is a special feature of high
$x$, i.e., low $W^2$, and does not imply the same is true for other regimes of $x$ and $Q^2$.

To conclude, by going to higher and higher orders in the leading-twist perturbative expansion we see that the only
strong evidence for higher-twist corrections is in the region of high $x$ and low $W^2$. All shortcomings of
low-order perturbative calculations seem to be reduced, if not removed, simply by working to higher orders.
Moreover, our particular knowledge of high $x$ coefficient functions leads us to believe that we have reached the
stage where the perturbative expansion and higher twist corrections are not really separable. This is further
illustrated by a method which goes beyond the standard logarithmic resummation using the `dressed gluon
exponentiation' approach of \cite{DGE}, which takes into account some all-order information on the kernel of
Sudakov resummation itself. In Ref.~\cite{GR}, this combined resummation of Sudakov logarithms, renormalon
contributions and higher twists was confronted by data on the Nachtmann moments (i.e. corrected for target mass)
of the structure function extracted from low $Q^2$ data. While it is impossible to estimate precisely the higher
twist contributions, one allowed description is that where the Sudakov resummation alone is necessary to explain
the high-moment data.

\section{Uncertainties due to input assumptions}
\label{sec:inputassumptions}

\subsection{Choice of input parameterization}

Over the years we have adopted parameterizations with more and more free parameters, as required by the increase
in precision and kinematic range of the data, and by the new types of data that have become available. Perhaps the
biggest single extension of the parameterization was the contribution added to the gluon that allowed it to
become negative at small $x$ \cite{MRSTNNLO1}. Recently other groups have investigated the influence of parton
parameterizations on parton uncertainties \cite{CTEQ6, H1new}, with various conclusions. In \cite{H1new}, which
describes a fit to a fairly limited set of data, it is found that there is no real improvement to the quality of
the fit once the number of parameters specifying the input parton distributions has increased past a certain
number, 10 in this case. In \cite{CTEQ6}, however, a new type of parton parameterization is used, and it is
claimed that this is necessary in order to obtain the best fit to the Tevatron jet data. Our studies do not
support this claim.

Although the fit to the Tevatron jet data in \cite{CTEQ6} is indeed better than in \cite{MRST2001}, and even in
\cite{MRST2002} (where the high $x$ gluon has been improved), we believe very little of this is due to the parton
parameterization. There are various differences between the CTEQ6 and MRST2002 approaches to global fits including
the choice of $Q^2$ cuts, the data sets included in the fit, the treatment of errors (particularly those of the
E605 Drell-Yan data) and even the definition of $\alpha_S(Q^2)$ at NLO. In \cite{MRSTkrakow} we discussed in
detail the effect of making our starting point for the fit more and more like that for CTEQ6, and discovered that
when we did so our fit to the jet data became of almost similar quality to that of CTEQ6 when using our own
parameterization (without any of the ``kinks'' found in some previous best fits to jet data \cite{MRST2001}). Some
of the further cuts we have introduced in the course of the investigations in this paper have led to even better
fits to the Tevatron jet data. Hence, we do not feel that these particular data require us to alter our parton
parameters to obtain the best possible fits. However, we do note that the fact that our parameterization does allow the 
input gluon to be negative does have important consequences compared to the CTEQ analysis. It means 
that the CTEQ6 gluon is always 
larger at very small $x$ than ours (even though their starting scale is a little larger, $Q_0^2=1.69~\GeV^2$,
our gluon is still negative at this $Q^2$), and from the momentum sum rule must be smaller elsewhere, in practice at 
intermediate $x$. This does lead to predictions which are significantly different to ours, examples being
the Higgs cross section at the Tevatron and LHC as seen in Fig.~15 of \cite{MRST2002}.    

Indeed, there is considerable evidence that we have, if anything, too much flexibility in the parton
parameterizations. When attempting to diagonalize the error matrix for partons when all the parameters were left
free we discovered that many were extremely correlated (or anti-correlated) leading to some very flat directions
in the eigenvalue space \cite{MRST2002}. In fact, in order to obtain a stable error matrix we found that it was
necessary to observe the departures away from the best fit while allowing only 15 of our nominal 24 parton
parameters to vary. A similar effect has, in fact, also been seen in \cite{CTEQHes}, where when producing the error
matrix only 16 out of a possible 22 parameters are allowed to vary, and in \cite{CTEQ6}, where only 20 out of 26
parameters are allowed to vary. Similar problems are not seen by other groups \cite{H1Krakow, H1new, ZEUSfit,
Alekhin}, all of whom use smaller data sets and a smaller number of parton parameters, implying that they have
either the correct number of parameters for the flexibility required by their data, or potentially slightly too
few.\footnote{As mentioned above, in \cite{H1new} the number of required parameters is carefully investigated. In
\cite{Alekhin}, however, one can check that more than 2\,$\sigma$ variations in the parton parameters would lead to
some pathological behaviour (particularly for the gluon), implying that the fit is on the verge of the same type
of redundancy problems as in \cite{MRST2002} and \cite{CTEQHes,CTEQ6}.}

The redundancy observed when calculating the error matrix
does not mean that all the other parameters can simply be dispensed with, since
it is necessary to allow most to vary in order to obtain the best fit itself,
and to allow the partons to have a sufficiently flexible shape. It just means that
not all the parameters need to vary in order to investigate small deviations from the
best fit partons. For example,
starting with our previous parameterization for the gluon
\be
xg(x,Q_0^2) = A_g(1-x)^{\eta_g}(1+\epsilon_g x^{0.5}+\gamma_g x)x^{\delta_g},
\ee
it was necessary to add a term of the form
\be
-A_-(1-x)^{\eta_-}x^{\delta_-},
\ee
in order to let the input gluon be negative at small $x$. Only 3 parameters were needed to investigate small
variations, but more parameters are certainly needed in order to obtain the gluon for the best fit. However, the
total of 7 free parameters we have for the gluon ($A_g$ is fixed by the momentum sum rule) do have some genuine
degree of redundancy. We have noticed in the course of performing many fits that, for any fixed value of
$\epsilon_g$ from $-1$ to $3$, we can obtain optimum descriptions of the data which are practically identical in
quality. In these fits all gluon parameters are significantly different, but the gluon distributions produced are
practically identical (at least between $x=0.9$ and $x=0.00001$). This is perhaps the clearest example of a
single parameter which is essentially redundant, but we have other similar examples.

Hence, we conclude that our input parameterizations are sufficiently flexible for the present data. We do not seem
to have the optimum parameterization for both finding the best fit and also investigating fluctuations about this
best fit. For a fully {\em global} fit, however, no-one else seems to have it either. To achieve this would require
fewer parameters than at present. This might then influence our error analysis using the Hessian approach somewhat,
but we feel it is unlikely to affect our best fit partons very much at all.

\subsection{Choice of heavy target corrections}

\begin{figure}[htbp]
\epsfig{figure=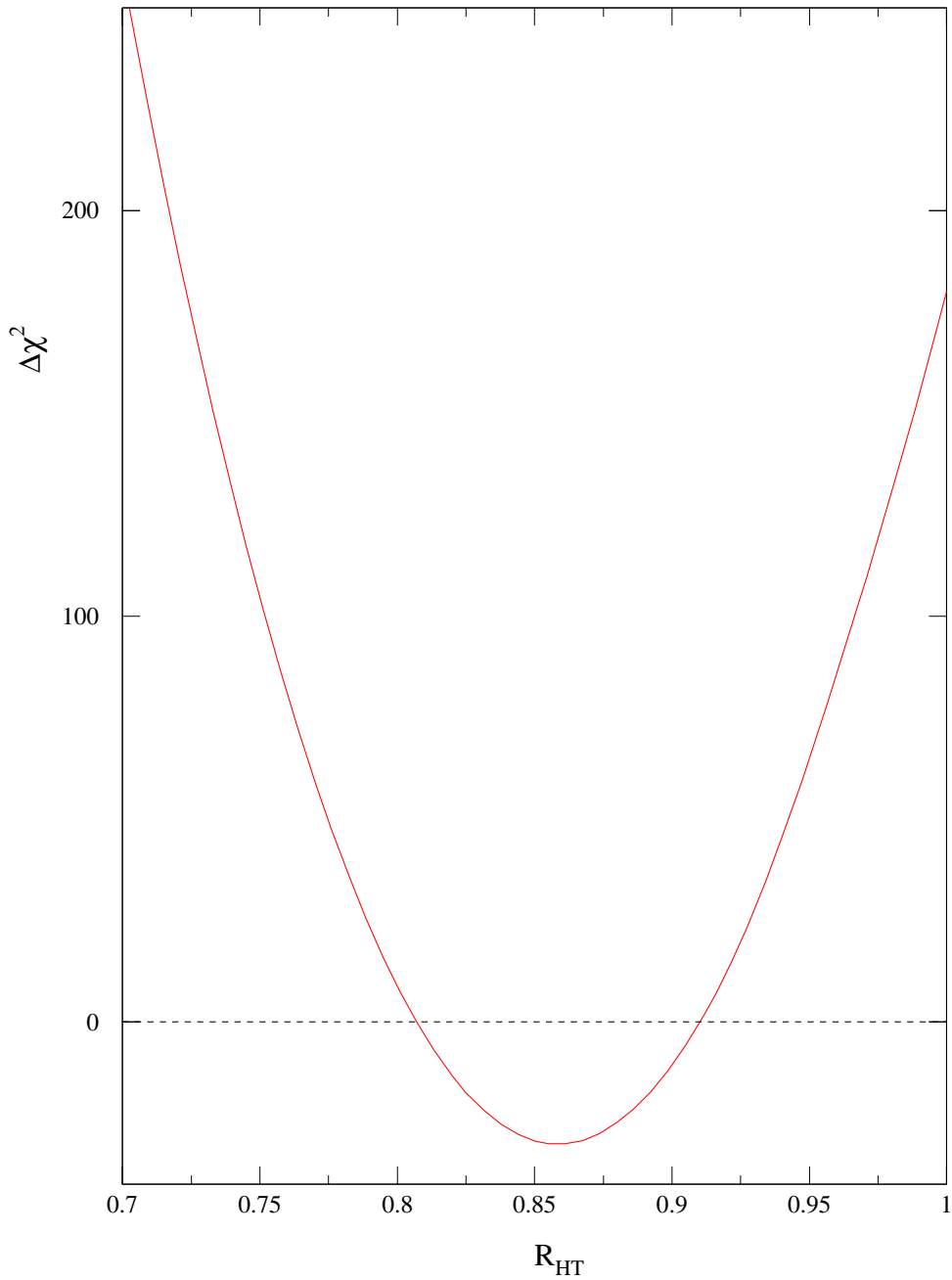,height=20cm}  
 \caption{\label{fig:f5a} The variation of $\chi^2$ for global fits with different heavy target corrections to the CCFR neutrino data.
 $\chi^2$ is plotted against the value of the heavy target correction $R_{\rm HT}$ at $x=0.0075$, as explained in the text.
 The default fit has $R_{\rm HT} = 0.807$, whereas for the optimum fit the correction is slightly less, $R_{\rm HT} = 0.86$.}
\end{figure}

\begin{figure}[htbp]
\epsfig{figure=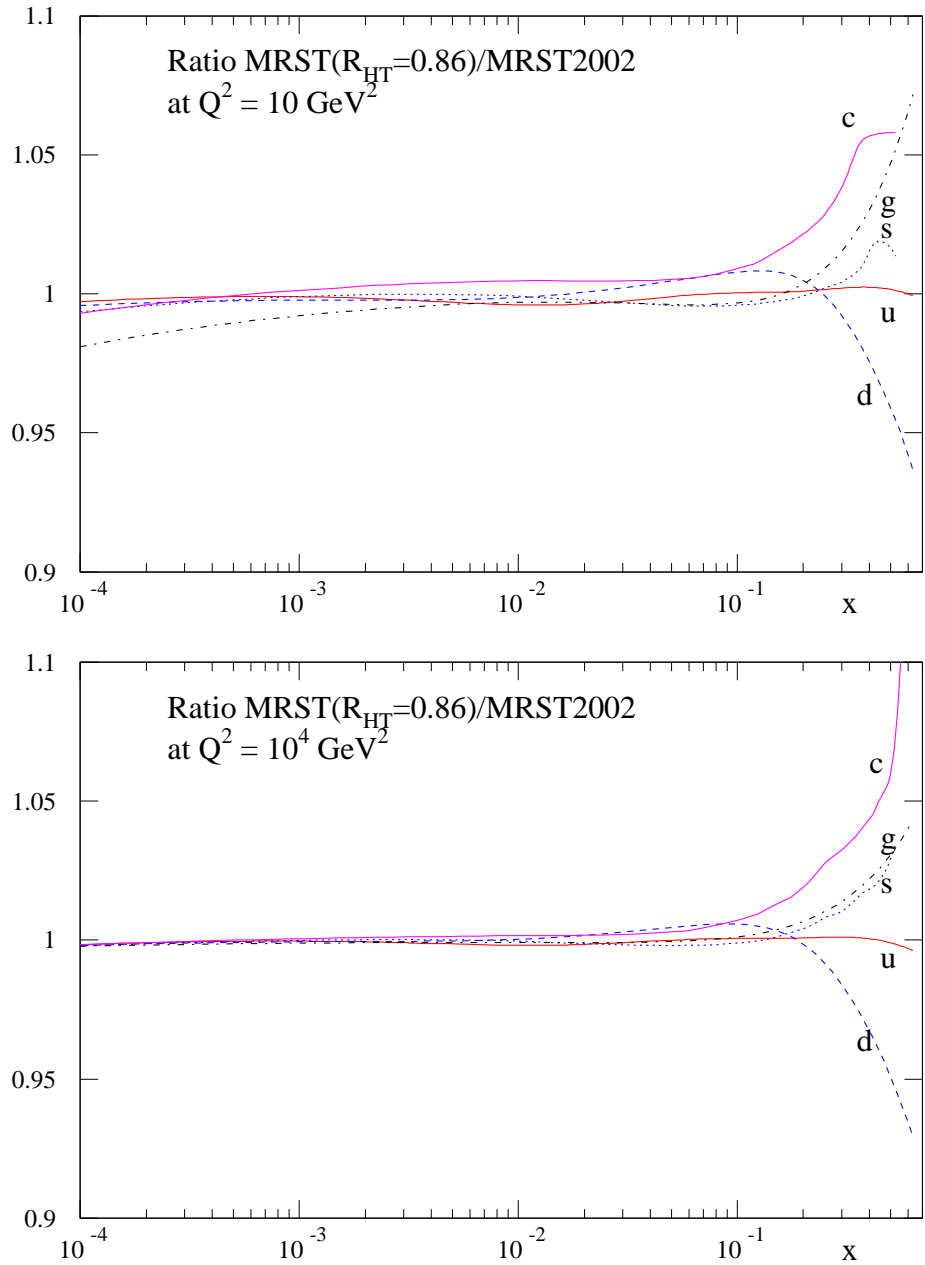,height=20cm}  
 \caption{\label{fig:htpart} The parton distributions obtained from the fit with the optimum shadowing correction of
 $R_{\rm HT}(x=0.0075)$ of $0.86$ compared with the default MRST2002 partons.}
\end{figure}

When fitting the CCFR $F^{\nu}_2(x,Q^2)$ and  $F^{\nu}_3(x,Q^2)$ data~\cite{CCFR} we have to use some model for
nuclear shadowing corrections. The form that we use for the heavy target correction factor is deduced from a
$Q^2$-independent fit to the EMC effect for the scattering of muons on a heavy nuclear target (A=56). To be
explicit, we parameterize the correction as
\be R_{\rm HT}\ =\ \left\{ \begin{array}{l l} 1.238 + 0.203\log_{10}x & {\rm for}\ x<0.0903\\
1.026 & {\rm for}\ 0.0903<x<0.234\\
0.783 - 0.385\log_{10}x & {\rm for}\ 0.234<x. \end{array}\right. \ee
In order to investigate the uncertainty arising from this correction we perform a series of fits, maintaining the
central plateau in $R_{\rm HT}$, but changing the slopes in $\ln x$ in the high and low $x$ regions. In addition,
we allow the normalisation to vary within the experimental error, as usual. We study the quality of the fit as a
function of the value of the heavy target correction at the lowest $x$ value ($x=0.0075$), for which CCFR data
exist. This data point receives the largest heavy target correction, that is $R_{\rm HT} = 0.807$ for our standard
fit. This is a very similar, but larger, correction to that for the highest $x$ value ($x=0.75$) for which data
exist.  Fig.~\ref{fig:f5a} shows the global $\chi^2$ as a function of the value of $R_{\rm HT}$ at $x=0.0075$.
Clearly our standard shadowing assumption, although near to the minimum, is not the absolutely optimum choice of
shadowing correction. The data prefers a value of  $R_{\rm HT}(x=0.0075)$ of $0.86$, i.e. a slightly smaller
correction than our usual choice, and the improvement in the fit is about $30$ units of $\chi^2$. Most of this
improvement comes from the fit to the CCFR data on $F_2^{\nu(\bar\nu)}(x,Q^2)$, which has always had a tendency to
lie underneath the theory when shadowing corrections are applied (see for example Fig.~9 of \cite{MRST2001}). The main
effect of the slightly reduced shadowing is just to bring these data in line with the theory (although there are
some other minor improvements) and the parton distributions themselves are not changed much at all, as shown in
Fig.~\ref{fig:htpart}.\footnote{The preliminary measurements of $F_2^{\nu(\bar\nu)}(x,Q^2)$ by the NuTeV
collaboration actually tend to lie a little above those of CCFR in the lowest $x$ bins , and hence lie rather
closer to the default theory curves with our standard shadowing correction \cite{NuTeVF2}.}

\bigskip

Similarly the deuterium data is corrected for shadowing effects. These prescriptions are not
unique. In our fits we normally use the parameterization in \cite{DEUTSHAD} which uses a theoretical model to 
estimate the nuclear shadowing corrections at relatively small $x$. There are other alternatives for 
models of this type of shadowing. There has also been a prescription for deuterium shadowing corrections 
extracted by the SLAC E139/140 experiments \cite{E139}. This is an empirical extraction which uses a model
\cite{Frankfurt} in which binding effects are assumed to scale with nuclear density. This gives a relatively large
shadowing correction for deuterium, especially at large $x$, and was used as a basis for an analysis of the
$d/u$ ratio in \cite{bodekyang}. The validity of this extraction is rather controversial (see e.g. \cite{thomas}),
but we use the results simply as an estimate on the uncertainty of the deuterium shadowing corrections. As such, 
a fit with this shadowing correction applied to the deuterium data gives an estimate of the model uncertainty from 
this source on the high $x$ valence partons, particularly $d(x,Q^2)$.\footnote{We note that an examination of
theoretical uncertainties due to nuclear effects in the deuteron has recently been studied in \cite{akl} 
with the aim of examining isospin depedence of higher twist corrections.}   
\begin{figure}[htbp]
\epsfig{figure=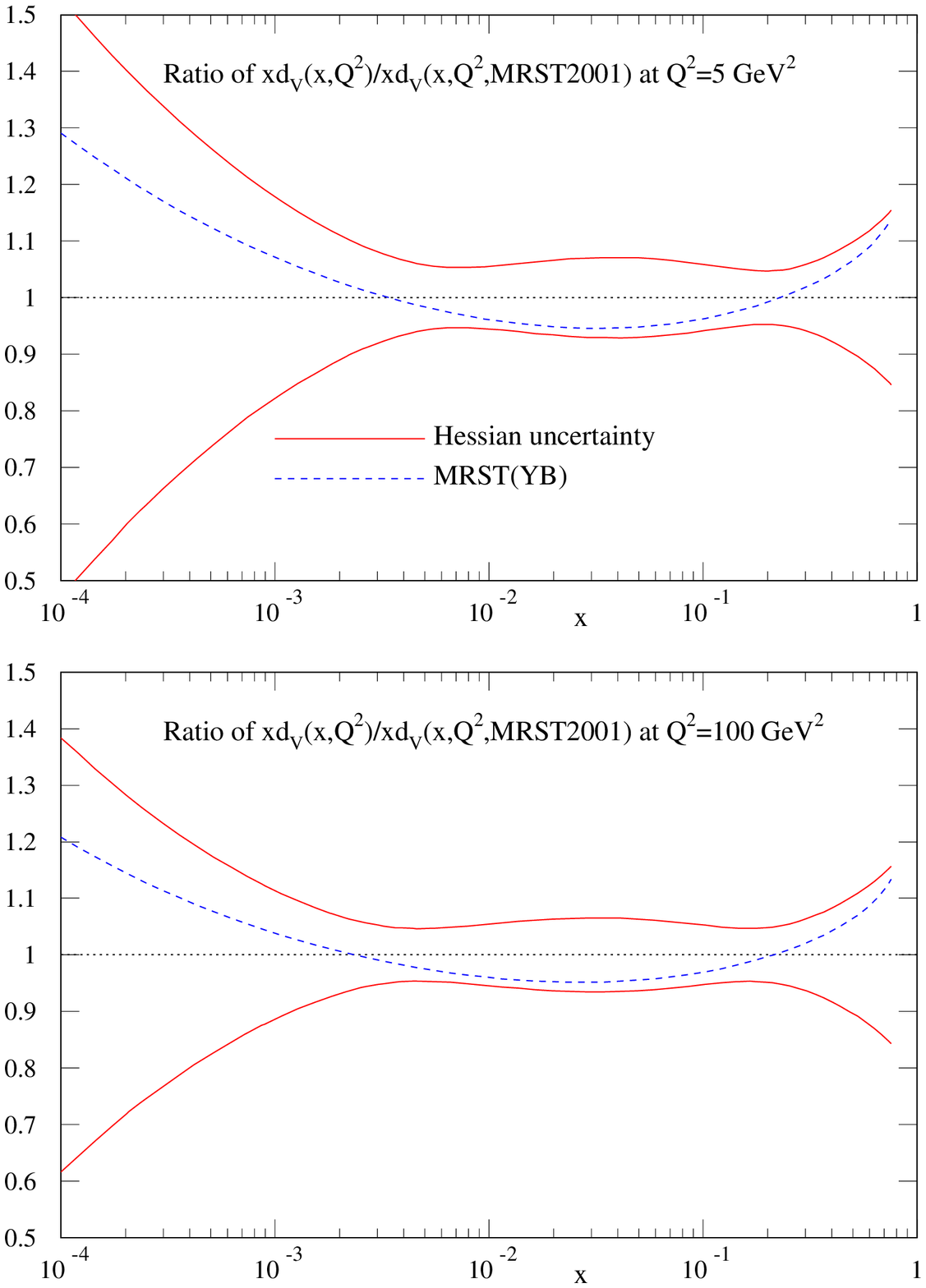,height=20cm}  
 \caption{\label{fig:deutshaddv} The ratio of $d_{\rm v}(x,Q^2)$ 
obtained from the fit with the deuterium shadowing correction of \cite{E139}
compared with the default MRST2002 partons, and also with the uncertainty in the default $d_{\rm v}(x,Q^2)$ 
distribution
from errors on experimental data found in \cite{MRST2002}.}
\end{figure}
\begin{figure}[htbp]
\epsfig{figure=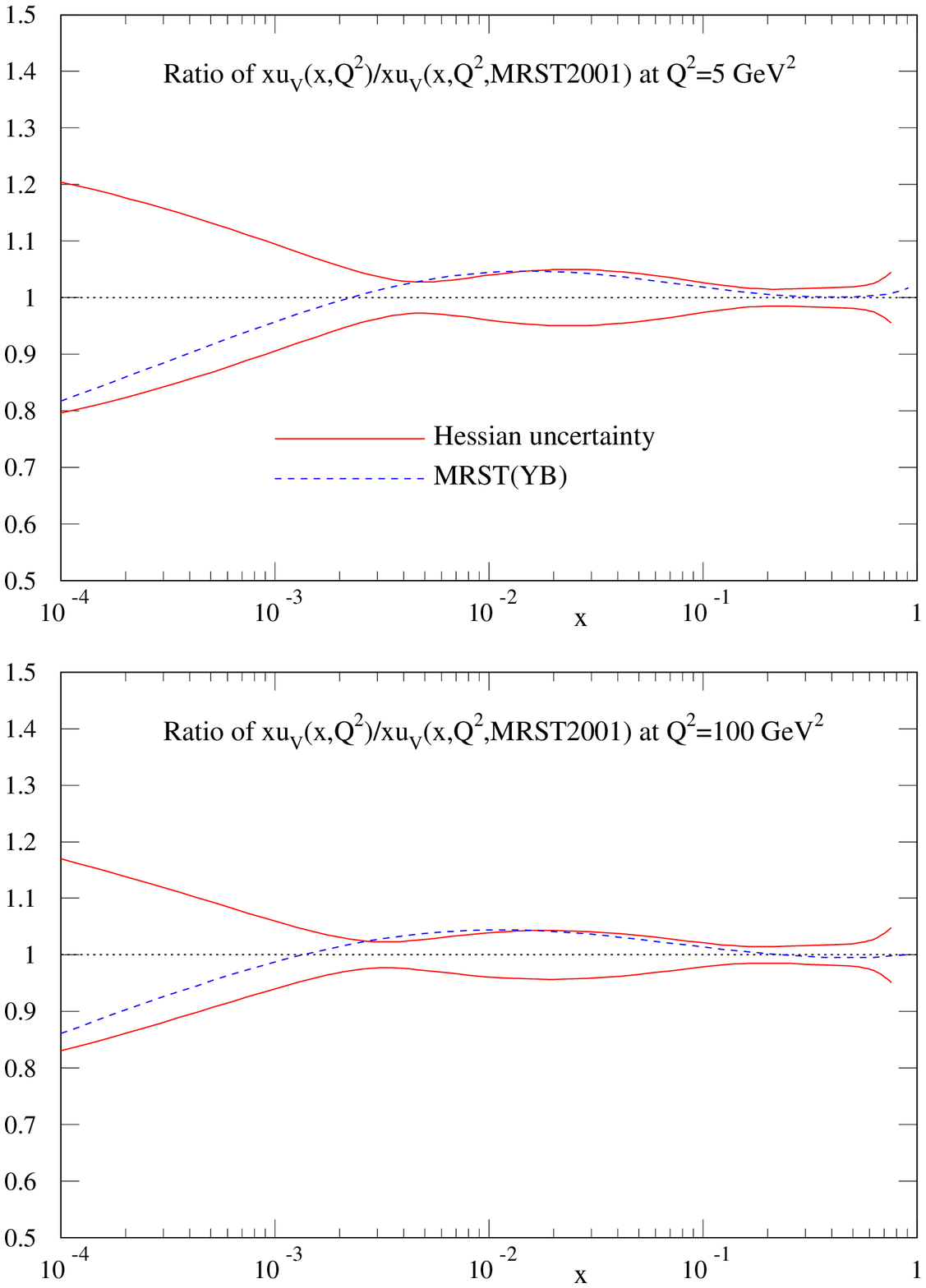,height=20cm}  
 \caption{\label{fig:deutshaduv} The ratio of $u_{\rm v}(x,Q^2)$ 
obtained from the fit with the deuterium shadowing correction of \cite{E139}
compared with the default MRST2002 partons, and also with the uncertainty in the default $u_{\rm v}(x,Q^2)$ 
distribution
from errors on experimental data found in \cite{MRST2002}.}
\end{figure}

This particular correction leads to a larger neutron structure function at high $x$. Hence, the largest change in 
the parton distributions is in the high $x$
down quark distribution, which increases significantly. As seen in Fig.~\ref{fig:deutshaddv} this increase is a little 
smaller at high $x$ than the uncertainty due to errors on experimental data that was estimated using 
the Hessian matrix method in \cite{MRST2002}. From the constraint on the number of valence quarks the increase in 
$d_{\rm v}(x,Q^2)$ at high $x$ must be compensated for elsewhere, and indeed we see that it is smaller than the
default for $x=0.01-0.1$, but is well within the bounds of the experimental uncertainty. The change in the 
$u_{\rm v}(x,Q^2)$ distribution must lead to a modification of the $u_{\rm v}(x,Q^2)$ distribution in order to obtain
the best overall global fit. This modification is shown in  Fig.~\ref{fig:deutshaduv}. Although it is proportionally
much smaller than the change in $d_{\rm v}(x,Q^2)$, $u_{\rm v}(x,Q^2)$ is much better constrained by data, and 
the change due to the different shadowing correction can be as big as, or even slightly larger than the 
uncertainty due to experimental errors on data. However, this is mainly so at $x<0.1$ where the direct constraint
on valence quarks is quite small since sea quarks dominate, and the estimated uncertainty from the Hessian method
for valence quarks alone is likely to have limitations related to parameterizations.    
Hence, for the down and up quark distributions we conclude that model errors on shadowing corrections in deuterium are
typically of the same size as the errors due to the experimental errors on deuterium structure function data.   

Finally we note that the quality of the global fit when using the deuterium corrections in \cite{E139} 
is 12 units in $\chi^2$, better than the default fit. There is a slight improvement in the total fit to 
deuterium data, mainly to the BCDMS data, but a slight deterioration in the fit to E605 Drell-Yan data. There is also 
a slight improvement in the fit to high-$E_T$ Tevatron jet data. This is to be expected since the increased down quark
distribution at high $x$ leads to an increase in the highest $E_T$ jet cross-section, which is what is required. 
However, the improvement is limited since an increase in $d(x,Q^2)$ at high $x$ increases the momentum carried by
the down quark. From the momentum sum rule this makes it harder to have a large gluon distribution at high $x$,
which the fit would like. In practice a compromise is reached, i.e. an even greater enhancement in $d_{\rm v}(x,Q^2)$
at high $x$ would slightly improve the fit to deuterium, and so other data, but actually makes the fit to Tevatron
jet data worse. The relatively small improvement in the global $\chi^2$ from this model of deuterium shadowing, 
partially limited by tension between different data sets and parton distributions, leads us to conclude that there is 
no strong supporting evidence for the model. Nevertheless, the small pull of the evidence is in favour of some
deuterium shadowing at high $x$.

\subsection{Size of input strange sea}

\begin{figure}[htbp]
\begin{center}
\epsfig{figure=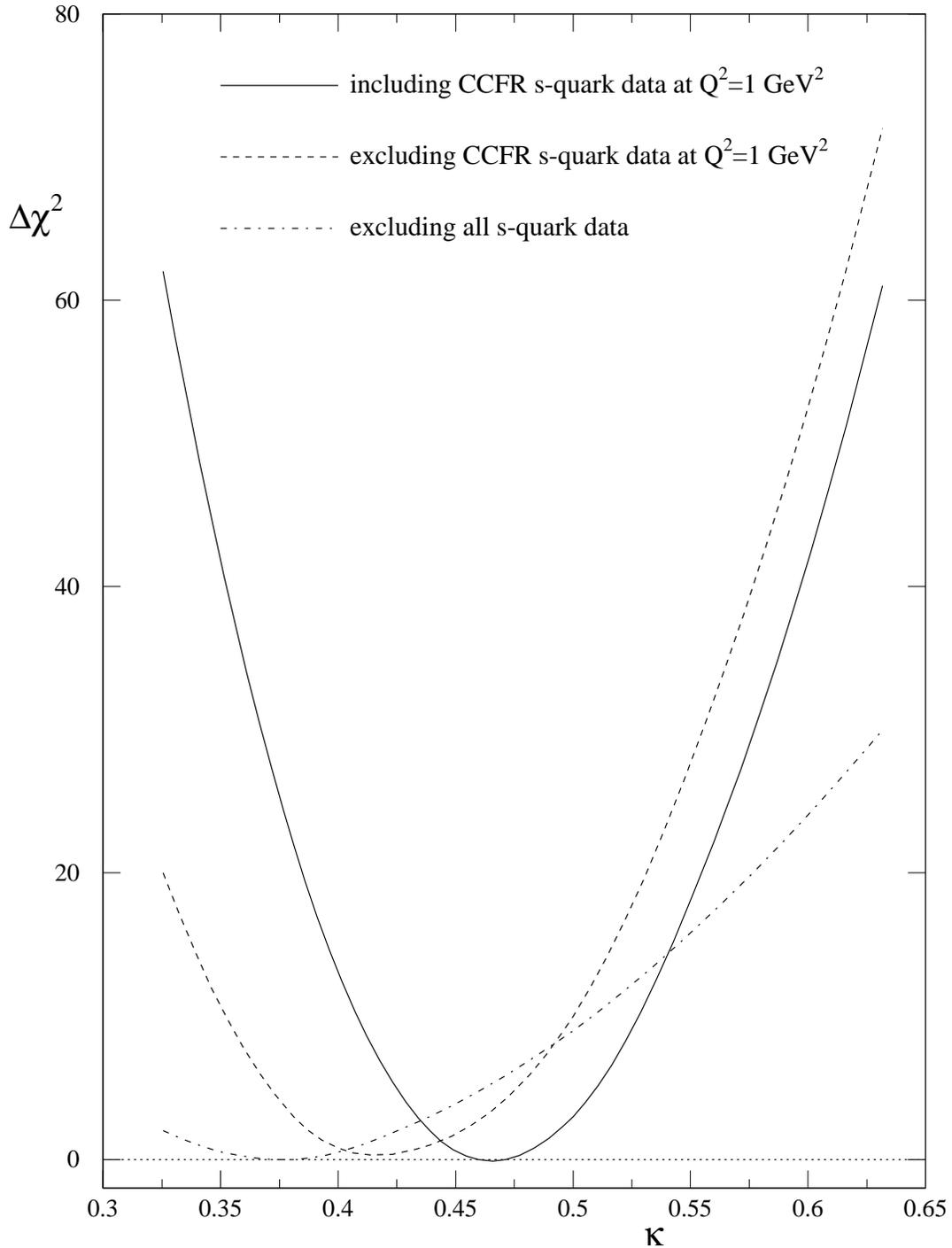,height=20cm}  
 \caption{\label{fig:f6} The variation of $\chi^2$ for global fits with different values of $\kappa$, defined in
 (\ref{eq:s}), which gives the strength of the strange quark sea relative to the non-strange sea. The two curves with
 the deep minima include the CCFR dimuon data~\cite{CCFRmm} in the global $\chi^2$, whereas when these data are
 omitted the shallower $\chi^2$ profile is obtained.}
\end{center}
\end{figure}

Global fits have traditionally assumed that the shape of the input strange quark sea distribution is the same as
the average of the input $\bar u + \bar d$ distribution. The primary experimental constraint on the strange distribution
is provided by data on dimuon production in neutrino--nuclei deep inelastic scattering~\cite{CCFRmm,NuTeV}. These
data are consistent with the shape assumption, and in addition constrain the magnitude of the strange distribution
to be approximately half of $(\bar u + \bar d)/2$ at low $Q^2$. For this reason, we conventionally choose
\be s(x) = \frac{1}{4}(\bar u(x) + \bar d(x)) \ee
at $Q_0^2 = 1~\GeV^2$. We now investigate the sensitivity to this input assumption.

First, we check the validity of our choice by expressing
\be s(x) = \kappa (\bar u(x) + \bar d(x))/2 \label{eq:s} \ee
and performing global fits, including the `data' on the strange sea provided by CCFR~\cite{CCFRmm}, for various
different fixed values of the parameter $\kappa$. The variation in $\chi^2$ is shown as a function of $\kappa$ in
%
%
Fig.~\ref{fig:f6}. The continuous and dashed curves correspond to including all CCFR data or omitting those CCFR
data at the lowest $Q^2$ value, $Q^2 = 1~\GeV^2$. The latter may be more appropriate since NLO leading-twist
distributions might become unreliable at such a low scale. We see that our default choice of $\kappa = 0.5$ is
near the minimum when the $Q^2 = 1~\GeV^2$ CCFR data are included, but is perhaps a little large when these data
are omitted. The choice $\kappa = 0.44$ appears optimum for both data selections. We therefore make a set of
partons available with this value. In the future, NuTeV dimuon data will better determine the strange sea.

We also investigate the effect of the variation of $\kappa$ on the quality of our standard global fit (in which
the contribution of the dimuon data is not included). This is shown by the (shallower) dash-dotted curve in
%
%
Fig.~\ref{fig:f6}. We see that the minimum is obtained for $\kappa = 0.38$ and that the $\chi^2$ is about 8 lower
than for the MRST(2002) set. Hence this is a rather small improvement in the fit and leads to a very small change
in the parton distributions. This value of $\kappa$ is near the limit of acceptability as determined by the dimuon
data. We conclude that the level of uncertainty of parton distributions, and related quantities, is rather small,
particularly as it weakened by evolution to higher $Q^2$.

\subsection{Possible isospin violations and $s\neq\bar s$}

Isospin symmetry implies that the parton distributions of a neutron are obtained from those of the proton simply
by swapping the up and down quark distributions, i.e. $d^n(x,Q^2) =u^p(x,Q^2)$ and $u^n(x,Q^2) =d^p(x,Q^2)$. In
the absence of any obvious evidence to the contrary this is always assumed to be true in global fits. There are
many sets of data in the global fit which would in principle be sensitive to any isospin violation. These are the various 
sets of deuterium structure
function data (SLAC, BCDMS, NMC), the CCFR neutrino structure function data from isoscalar targets, the E605
Drell--Yan data on a copper target, and the NA51 and E866 Drell--Yan asymmetry\footnote{The $pp$ and $pn$
asymmetry measurements involve Drell--Yan production on deuterium, as well as hydrogen, targets.} data. However, in the
global fit, all of these data sets are well described, implying that isospin symmetry breaking is small. This may
be illustrated as follows. Assuming that the structure functions are dominated by the up and down quarks (and
antiquarks), which is largely true, then the measurements of the $\gamma$-exchange contributions to $F_2$
determine the quark contributions
\be F_2^p \propto \frac{4}{9}\left(u^p + \bar u^p\right) + \frac{1}{9}\left(d^p+ \bar d^p\right) \label{eq:F2p}\ee
\be F_2^n \propto \frac{4}{9}\left(u^n + \bar u^n\right) + \frac{1}{9}\left(d^n+ \bar d^n\right) \label{eq:F2n}\ee
whilst the structure functions for $\nu$ and $\bar\nu$ interactions on an isoscalar target constrain the
combinations
\be F_2^\nu \propto d^p + \bar u^p + d^n+ \bar u^n \label{eq:F2nu}\ee
\be F_2^{\bar\nu} \propto u^p + \bar d^p + u^n+ \bar d^n \label{eq:F2nubar}\ee
\be xF_3^\nu \propto d^p - \bar u^p + d^n- \bar u^n \label{eq:F3nu}\ee
\be xF_3^{\bar\nu} \propto u^p - \bar d^p + u^n- \bar d^n. \label{eq:F3nubar}\ee
However the CCFR data~\cite{CCFR} for the neutrino structure functions $F_2$ and $xF_3$ that are used in the
global analyses average over the $\nu$ and $\bar\nu$ interactions\footnote{When available, the NuTeV measurements
of the $\nu$ and $\bar\nu$ interactions individually will offer even more stringent checks of the isospin
relations.}. Note that the $F_2^n/F_2^p$ NMC data determine the ratio of the quark combinations of (\ref{eq:F2n})
and (\ref{eq:F2p}) with about a 2\% error. The data thus allow little flexibility in the neutron parton
distributions even without the additional constraints of Drell--Yan data and $W$-asymmetry data (which constrain
the proton valence and sea quarks).

\begin{figure}[htbp]
\epsfig{figure=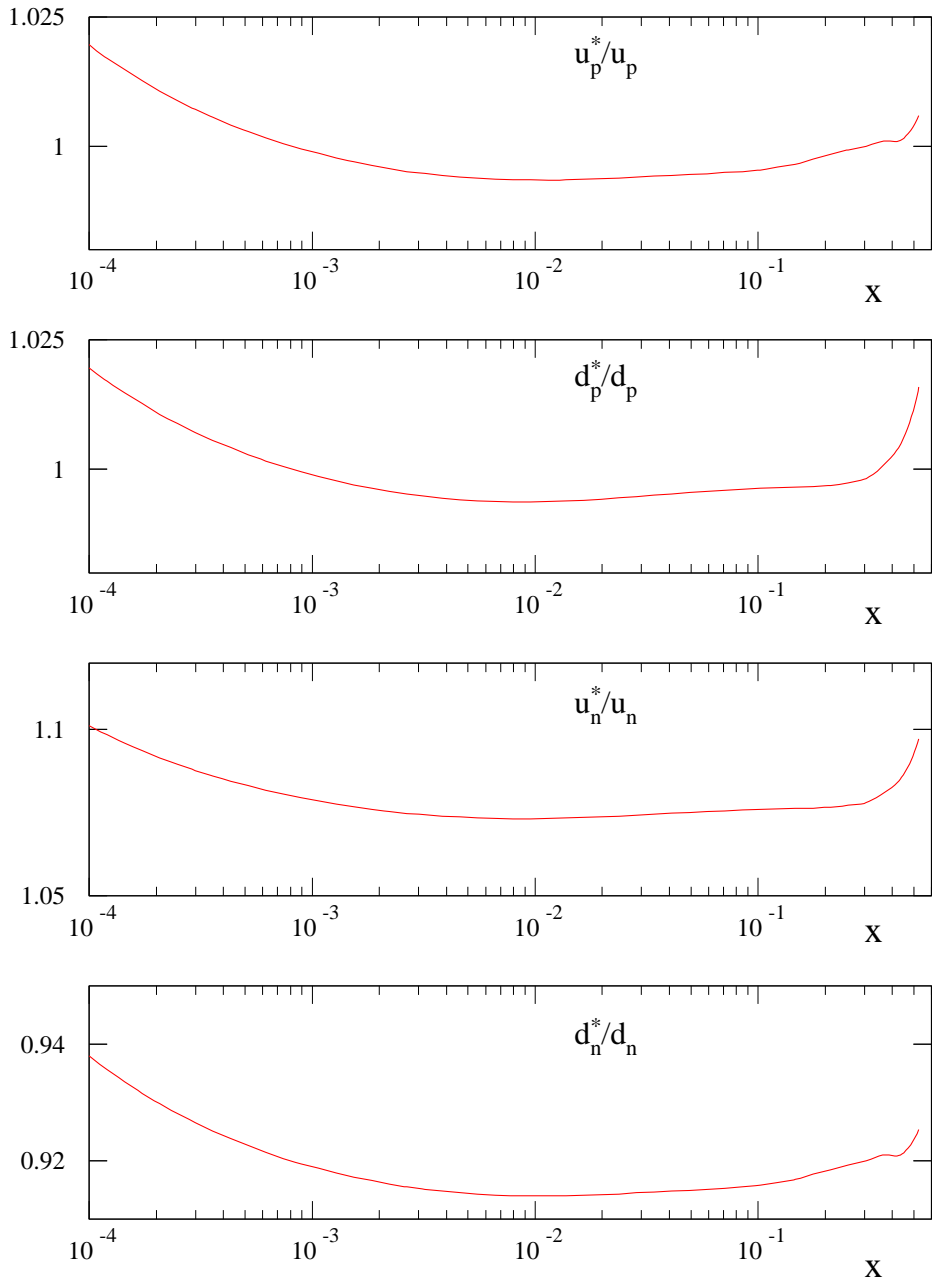,height=20cm}       
 \caption{\label{isosea} The up and down sea quarks for the best fit with allowed isospin violation in the sea quarks
compared to the default partons. The ratio is shown both for the proton and neutron sea quark distributions.}
 \end{figure}

We consider the possibility of isospin violation in the valence quarks and the sea quarks separately. For the sea
quarks we assume
\be u_{\rm sea}^n(x) = d_{\rm sea}^p(x)(1 +\delta), \label{eq:u_sea} \ee
\be d_{\rm sea}^n(x) = u_{\rm sea}^p(x) (1 -\delta). \label{eq:d_sea} \ee
This type of violation is expected from theoretical models, and is consistent with momentum conservation, up to
very small violations due to a non-zero value of $(d^p_{\rm sea} - u^p_{\rm sea})\delta$. Strictly speaking it is
not preserved by evolution of the partons, but in the kinematic regions of interest the violation is very small. Somewhat
surprisingly, we find that a certain amount of isospin violation of this type is actually preferred by the data.
The best fit is obtained for $\delta = 0.08$, i.e. an $8\%$ violation of isospin in the sea. This fit has a total
$\chi^2$ 20 better than the default best fit at NLO. The majority of this improvement comes from the fit to NMC
data on $F^n_2/F^p_2$, which is raised a little by the increase in $u_{\rm sea}^n(x)$. The fit to the E605
Drell--Yan data is also markedly improved. The change in the up and down sea quarks for this fit compared
to our default partons is shown in Fig.~\ref{isosea}. This clearly shows the preference of the data for the up
sea distribution in the neutron to be enhanced, and the down quark suppressed.

\begin{figure}[htbp]
\epsfig{figure=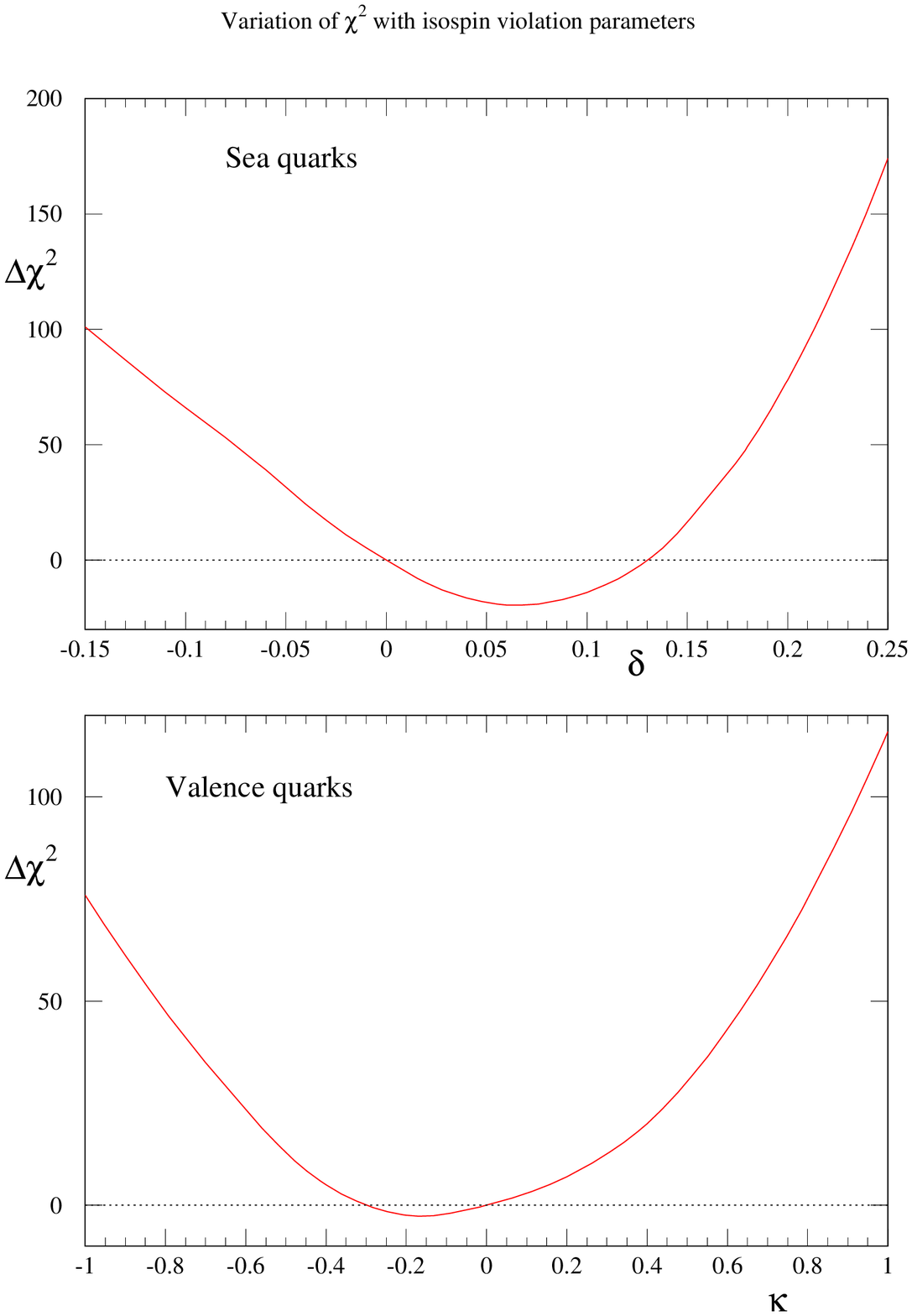,height=20cm}       
 \caption{\label{isospin} $\chi^2$ profiles showing the effect of isospin violation in the sea and valence
 quark sectors respectively. $\delta$ and $\kappa$ are defined in eqs.~(\ref{eq:u_sea}) 
and (\ref{eq:u^n_v}) respectively.}
 \end{figure}

An increase in $\chi^2$ of 50, corresponding to a $90\%$ confidence
limit from our arguments in \cite{MRST2002}, arises when $\delta=0.18$ or $\delta = -0.08$, see
Fig.~\ref{isospin}. In the former case the NMC data on $F^n_2/F^p_2$ (which is the most sensitive discriminator
of isospin violation) has now become too large, the fit to BCDMS $F_2^p$ data has deteriorated\footnote{The
deterioration is caused by the decrease of $d_{\rm sea}^p(x)$, which arises in order to prevent $u_{\rm sea}^n(x)$
becoming too large.} and the description of the E866 Drell--Yan asymmetry has become very poor. The fit to E605
data has continued to improve, however. For $\delta=-0.08$ the prediction for $F^n_2/F^p_2$ has become too small,
and the fit to E605 Drell--Yan data has deteriorated.

For the valence quarks a similar model of isospin violation is not possible because it would violate the valence
quark number counting for the neutron. Hence we consider a violation of the type
\be u^n_{\rm v}(x) = d^p_{\rm v}(x) + \kappa f(x),\label{eq:u^n_v}\ee
where $f(x)$ is a function which has zero first moment. A suitable function, which has the same type of behaviour
as the valence quarks at high and low $x$ is $f(x) = (1-x)^4x^{-0.5}(x-0.0909)$. However, there is a further
constraint. If we add $\kappa f(x)$ in order to obtain $u^n_{\rm v}(x)$, then we must subtract $\kappa f(x)$ in
order to obtain $d^n_{\rm v}(x)$, i.e.
\be d^n_{\rm v}(x) = u^p_{\rm v}(x) - \kappa f(x),\label{eq:d^n_v}\ee
in order to ensure momentum conservation. Hence, we consider isospin violation of this form. Again, the isospin
violation is not exactly preserved by evolution, but is correct to a very good approximation.

For valence quarks, there is very little preference for isospin violation. The best fit is obtained for
$\kappa=-0.2$, see Fig.~\ref{isospin}, but this gives an improvement in $\chi^2$ of only 4. It corresponds to
at maximum about a $3\%$ violation of isospin for $u^n_{\rm v}(x)$. 
The $90\%$ confidence level is obtained for $\kappa =
-0.8$ or $\kappa= 0.65$. In the former case, the main deterioration is in the description of the CCFR
$F_2^{\nu}(x,Q^2)$ data in the region of $x=0.2$, and in the latter case, it is CCFR $F_3^{\nu}(x,Q^2)$ data and
BCDMS $F_2^{d}(x,Q^2)$ data which are both badly fit at $x \sim 0.5$. For positive $\kappa$, $d^p_{\rm v}(x)$
decreases at high $x$ to compensate the isospin violating term and in order to fit the $F^p_2(x,Q^2)$ data
$u^p_{\rm v}(x)$ decreases, but less severely (due to the higher charge weighting). Hence, the failure in the fit
occurs when $u^n_{\rm v}(x)$ has increased too much for the BCDMS deuterium data but the larger decrease in
$d^p_{\rm v}(x)$, compared to the increase in $u^p_{\rm v}(x)$, leaves the prediction for $F_3^{\nu}(x,Q^2)$ (which is
sensitive to $u^p_{\rm v}(x)+d^p_{\rm v}(x)$) too small. For negative $\kappa$, $d^p_{\rm v}(x)$ increases at high
$x$, to compensate the isospin violating term, and in order to fit the $F^p_2(x,Q^2)$ data, $u^p_{\rm v}(x)$
increases less severely. In this case $u^p_{\rm v}(x)+d^p_{\rm v}(x)$ decreases, and is too small for
$F_2^{\nu}(x,Q^2)$. These upper and lower limits represent an isospin violation of at most ${\cal O}(10\%)$ for
$u^n_{\rm v}(x)$.  We offer no theoretical model for why our isospin violation is of the form seen, and note that 
calculations of the effect, e.g. \cite{Rodionov}, often indicate smaller results. However, these calculations are
very difficult, depending on intrinsically nonperturbative physics, and relying on models and assumptions. We simply
examine the empirical evidence given by the data.

It is interesting to compare the level of isospin violation with that needed to explain 
the NuTeV $\sin^2 \theta_W$ anomaly \cite{NuTevtheta}. The quantity measured\footnote{The NuTeV experiment does
not exactly measure $R^-$, in part because it is not possible experimentally to measure neutral
current reactions down to zero recoil energy, see \cite{NuTevtheta,NuTeVtheta2}.} by NuTeV is
\be R^-=\frac{\sigma^{\nu}_{\rm NC}
-\sigma^{\bar\nu}_{\rm NC}}{\sigma^{\nu}_{\rm CC}
-\sigma^{\bar\nu}_{\rm CC}}. \label{eq:rminus}\ee
In the simplest approximation, i.e. assuming an isoscalar target, 
no isospin violation and equal strange and anti-strange distributions, this ratio is given by
\be R^-\approx \frac{1}{2}-\sin^2 \theta_W,\label{eq:rminusn}\ee 
and so the measurement gives a determination of $\sin^2 \theta_W$. NuTeV find $\sin^2 \theta_W=0.2277 \pm 
0.0013({\rm stat.}) \pm 0.0009({\rm syst.})$ \cite{NuTevtheta}, compared to the global average of $0.2227 \pm 0.0004$,
i.e. about a $3\sigma$ discrepancy. However, if one allows for isospin violation then the simple expression
becomes modified to
\be R^-=\frac{1}{2}-\sin^2 \theta_W +(1-\frac{7}{3}
\sin^2 \theta_W) \frac{[\delta U_{\rm v}] -[\delta D_{\rm v}]}{2[V^-]}\label{eq:rminusi},\ee
where
\be [\delta U_{\rm v}] = \int_0^1\,x(u^p_{\rm v}(x) - d^n_{\rm v}(x)), \qquad\qquad  
[\delta D_{\rm v}] = \int_0^1\,x(d^p_{\rm v}(x) - u^n_{\rm v}(x)),\label{eq:momdefs}\ee  
are measures of the inequality in momentum fraction of the valence quarks 
induced by isospin violation, and $[V^-]\approx 0.45$ is the overall momentum fraction carried by the valence quarks.  
One can easily see that given a fixed value of measured $R^-$, a negative value of $\kappa$ 
moves the extracted value of $\sin^2 \theta_W$ downwards. The approximate effect can be found simply by using 
Eq.~(\ref{eq:rminusi}), but a more precise result is obtained by applying the functionals 
presented in \cite{NuTeVtheta2},
which account for the complications of the measurement. This reduces the naive modification by $\sim 10\%$. 
Using our best fit value of $\kappa=-0.2$ we obtain $[\delta U_{\rm v}]= -[\delta D_{\rm v}]=0.002$ 
(corresponding to $<0.5\%$ of the 
momentum carried by the valence quarks) and 
the modification is $\Delta\sin^2 \theta_W =-0.0018$. Hence, about 
$1\sigma - 1.5 \sigma$ of the $3\sigma$ discrepancy is removed. The determination of $\sin^2 \theta_W$ is far less 
sensitive to the isospin violation in the sea quarks. Our preferred violation with $\delta=0.08$ is in the wrong
direction to account for the discrepancy, but only reduces the effect from the valence quarks slightly to 
$\Delta \sin^2 \theta_W = -0.0015$.
Hence, the total result from the best fits with allowed 
isospin violation is to reduce the $3\sigma$ discrepancy to a $2\sigma$ 
discrepancy. However, the allowed range of isospin violation could easily allow the discrepancy to be 
removed altogether, or even to be made worse. It is nevertheless interesting that the weak indication given by 
the data in the global fit is 
such as to reduce the discrepancy by a significant amount.  

One can also investigate the possibility of $s(x) \not= \bar s(x)$. The only data in the fit which are sensitive
to this difference are the NuTeV dimuon data \cite{NuTeV}. These data are more difficult to analyse than the
similar CCFR data because they are presented in a much more exclusive form. However, in principle such data will
allow for a much more detailed analysis. The NuTeV group themselves have performed an analysis of their data
allowing the $s(x)$ and $\bar s(x)$ distributions to have different normalisations and different $(1-x)^\eta$
behaviour at high $x$. Their analysis indicates that the data would prefer a slight (11\%) excess of $\bar s(x)$
over $s(x)$ \cite{NuTeV}. However the analysis is at leading order, and does not impose the quark counting rule,
i.e. equal number of strange and anti-strange quarks. It is complicated to improve this analysis to the full NLO
level. CTEQ have performed similar preliminary analyses \cite{Olness}, obtaining very similar results. Most
predictions of physical quantities are insensitive to the potential imbalance of $s(x)$ and $\bar s(x)$, but the
NuTeV $\sin^2 \theta_W$ anomaly is affected by any difference~\cite{NuTevtheta}. However, the excess of $\bar
s(x)$ over $s(x)$ actually makes the discrepancy slightly worse. More precise data and a full theoretical
treatment will hopefully lead to an improved understanding of this question in future.\footnote{CTEQ have produced
fits where the quark counting is imposed and find that the excess in $\bar s(x)$
over $s(x)$ in the region of data must then be countered by an excess of $s(x)$ over $\bar s(x)$ at higher $x$,
leading to a positive momentum excess of $s(x)$ over $\bar s(x)$ \cite{Wu-Ki}. 
This is then in the correct direction to
reduce the NuTeV $\sin^2 \theta_W$ anomaly.}

\section{Implications for $\a$ and predictions for observables}
\label{sec:implications}

\subsection{Determination of $\a$}\label{sec:determination}

As we have already demonstrated, there is a significant amount of variation in our extracted value of 
$\alpha_S(M_Z^2)$ when we vary
the input assumptions for our fitting procedure, particularly when we vary the $x$ and $Q^2$ cuts. Previously we
have always determined  $\alpha_S(M_Z^2)$ from a global fit using our default cuts, that is no cut on
$x$ and a $Q^2$ cut of $2~\GeV^2$. As demonstrated earlier, the evidence suggests that when fitting this full
range of data, standard NLO (or NNLO) perturbation theory is not completely sufficient, and the variation in
$\alpha_S(M_Z^2)$ is due to the parameters in the fit compensating for the deficiencies of the theoretical
treatment. Hence, the `true' value of $\alpha_S(M_Z^2)$ should be that corresponding to the `conservative' partons
at both NLO and NNLO. We present these values below in  Table~\ref{tab:t4}, labelled MRST03, together with other
recent determinations.  Note that the conservative NLO partons themselves use $\alpha_S = 0.1162$. However the
$\chi^2$ profile versus $\alpha_S$ is very flat between 0.116 and 0.117. We therefore choose the mid-point as the
best value of $\alpha_S$, and $\Delta\chi^2=5$ to give the $1\sigma$ error of $\pm0.002$.

\begin{table}\begin{center}\begin{tabular}{l c l}
\hline\hline
\rule[-2ex]{0ex}{5ex} &$\Delta\chi^2$&$\alpha_S(M_Z^2)\ \pm{\rm expt}
\pm{\rm theory}\pm{\rm model}$\\
\hline
\rule[-2ex]{0ex}{5ex}\fbox{NLO}&&\cr CTEQ6&  100 & $0.1165\pm0.0065$ \cr ZEUS & 50 &
$0.1166\pm0.0049~~~~~~~~~~\pm0.0018$ \cr MRST03 & 5 & $0.1165\pm0.002~\,\pm0.003$ \cr H1 & 1  &
$0.115~\,\pm0.0017\pm0.005 ^{+0.0009}_{-0.0005}$ \cr  Alekhin & 1 & $0.1171\pm0.0015\pm0.0033$ \cr
\hline
\rule[-2ex]{0ex}{5ex}\fbox{NNLO} &&\cr MRST03 & 5 & $0.1153\pm0.002\pm0.003$
\cr Alekhin & 1 &
$0.1143\pm0.0014\pm0.0009$ \cr
\hline\hline
\end{tabular}
\caption{\label{tab:t4} The values of $\alpha_S(M_Z^2)$ found in NLO and NNLO fits to DIS data. The experimental
errors quoted correspond to an increase $\Delta\chi^2$ from the best fit value of $\chi^2$. CTEQ6~\cite{CTEQ6} and
MRST03 are global fits, where the latter correspond to the `conservative' sets of partons of
Sections~\ref{sec:effectcombined} and \ref{sec:cutsNNLO}. H1~\cite{H1Krakow} fit only a subset of $F_2^{ep}$ data,
while Alekhin~\cite{alekhin03} also includes $F_2^{ed}$ and ZEUS~\cite{ZEUSfit} in addition include $xF_3^\nu$
data.}
\end{center}\end{table}

As mentioned in Section~\ref{sec:selection}, a Bayesian approach to determining parton uncertainties~\cite{Giele}
gives $\alpha_S(M_Z^2) = 0.112\pm0.001$. However, in order to satisfy the strict requirements of consistency
between the data sets, this Bayesian analysis only uses the BCDMS~\cite{BCDMS}, E665~\cite{E665} and
H1(94)~\cite{H194} data for $F_2^p$. Bearing in mind that the H1(94) data have relatively large errors, the value
of $\alpha_S(M_Z^2)$ that is obtained simply reflects the original BCDMS determination of Ref.~\cite{MV}.

\begin{figure}[htbp]
\epsfig{figure=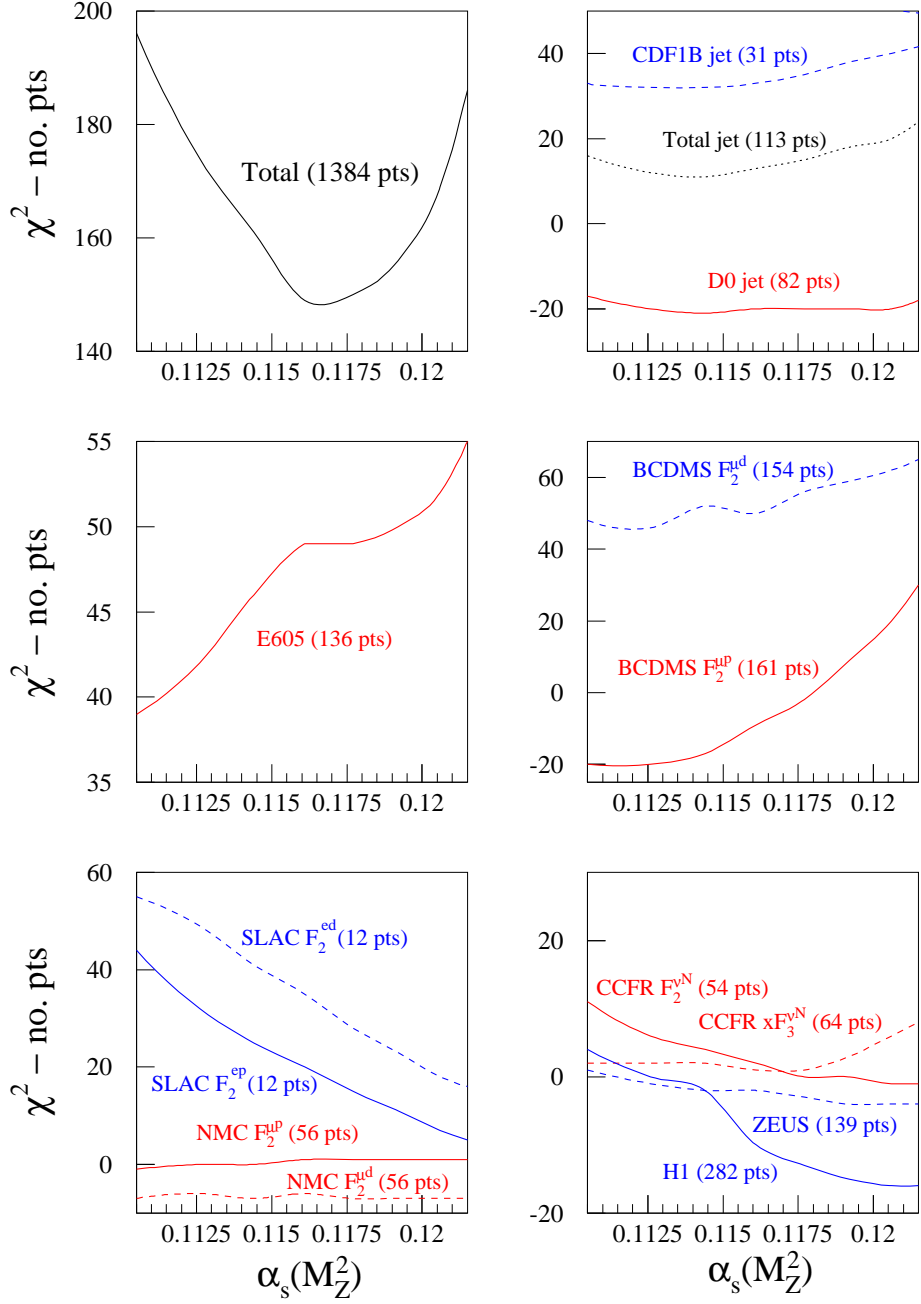,height=20cm}
 \caption{\label{fig:chisq16} The quality of the fit to the individual data sets included in the NLO global analysis
with $x_{\rm cut} = 0.005$ and $Q^2_{\rm cut} =10~\GeV^2$,
shown together with the overall total $\chi^2$, as a function of $\alpha_S(M_Z^2)$.}
\end{figure}

From Table~\ref{tab:t4} we see that the various determinations of $\alpha_S(M_Z^2)$ have approximately converged
to a common value, even though they are based on different selections of the DIS and related data. Averaging the
two global NLO analyses we have
\be \alpha_S(M_Z^2) = 0.1165\pm 0.004. \ee
Previously, the MRST value~\cite{MRST2001} was larger, $ \alpha_S(M_Z^2) = 0.119$. 
This was due to the attempt to fit data in regions
where the theoretical corrections to NLO DGLAP appear to be important. It is well illustrated by Fig.~16 of
\cite{MRST2001}, where we see that the optimum value of $\alpha_S(M_Z^2)$ varies considerably from data set to
data set. However, when the `conservative' data cuts are applied, the tension between the data sets is reduced
enormously, as can be seen by comparing Fig.~\ref{fig:chisq16}  with Fig.~16 of \cite{MRST2001}. Some data sets, 
particularly the SLAC data, which prefer a high value of $\alpha_S(M_Z^2)$, and the E605 Drell-Yan data and the 
BCDMS data, which prefer a low value of $\alpha_S(M_Z^2)$, still pull strongly away from the minimum value. 
However, the other data sets are now at, or near, their minimum $\chi^2$ for the best fit value 
of $\alpha_S(M_Z^2)$, which was certainly not the case for the default fit. The D0 and CDF jet data are 
a particularly good example, where not only are 
the data better fit by the `conservative' sets than the default sets, but they are no longer pulled to 
extreme values of $\alpha_S(M_Z^2)$ in order to obtain their best individual fits. 
The same marked improvement occurs for the NMC data which survive the $x$ and $Q^2$ cuts.
This increased compatibility between the data sets, and also between the data and the theory, 
is why the tolerance $\Delta\chi^2$ has been reduced from 20 (in~\cite{MRST2001}) to
5 in the present study. The reduction in tension between data sets also occurs when $\ln(1/x)$, $\ln(1-x)$ or higher-twist corrections 
are included, as discussed in Section~\ref{sec:specific}.

There are fewer extractions of the value of $\alpha_S(M_Z^2)$ using NNLO global or semi-global fits to DIS and
related data. The results are also shown in Table~\ref{tab:t4}. Again we see good agreement, and a small, but
definite, reduction from the NLO value. We found, in this case, far less sensitivity to the data cuts,
indicating that some important theoretical corrections are already accounted for at NNLO.

\subsection{Predictions for $W$ and Higgs hadroproduction}

\begin{figure}[htbp]
\begin{center}
\epsfig{figure=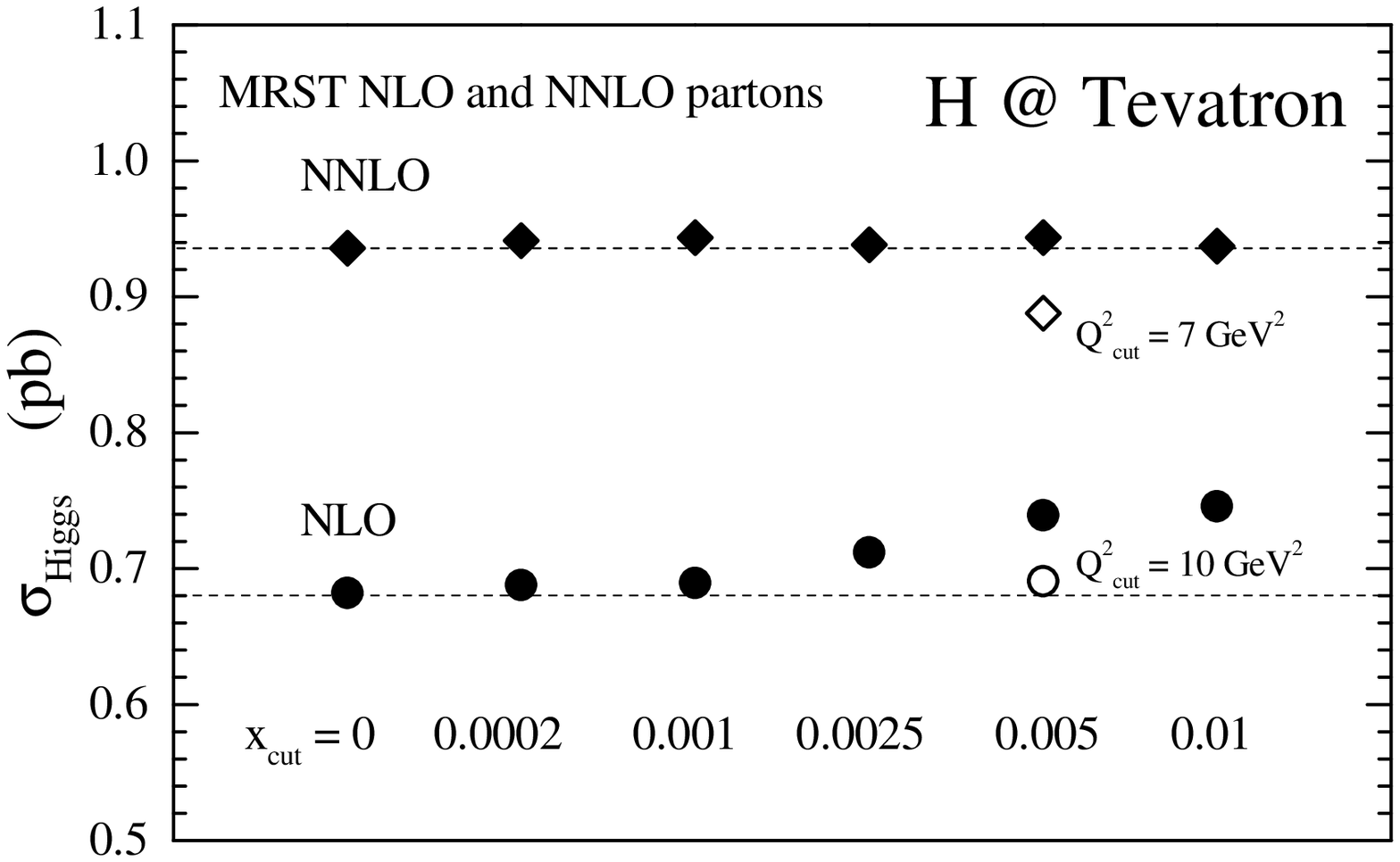,height=8cm}  
 \caption{\label{fig:HTev} Predictions for the $(M_H = 120~\GeV)$ Higgs cross section at the 
Tevatron ($\sqrt s = 1.96~\TeV$) at
 NLO and NNLO for various values of $x_{\rm cut}$, and for the `conservative' partons with a cut on both $x$ and
 $Q^2$ (shown as open symbols).}
\end{center}
\end{figure}

\begin{figure}[htbp]
\begin{center}
\epsfig{figure=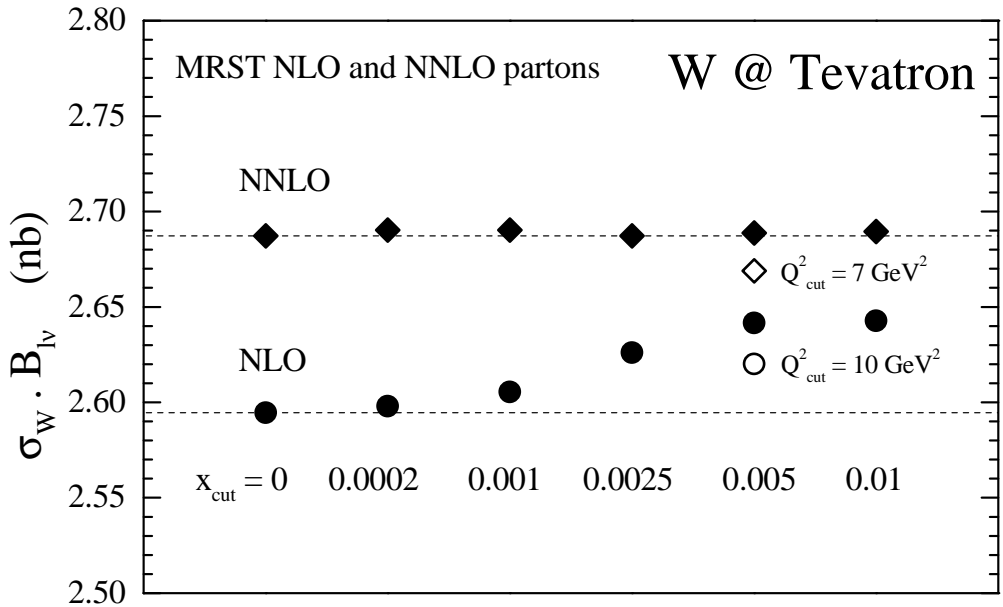,height=8cm}  
 \caption{\label{fig:WTev} Predictions for the W cross section (times the leptonic branching ratio $B_{l\nu}=0.1068$)
at the  Tevatron ($\sqrt s = 1.96~\TeV$) at
 NLO and NNLO for various values of $x_{\rm cut}$, and for the `conservative' partons with a cut on both $x$ and
 $Q^2$ (shown as open symbols).}
\end{center}
\end{figure}

\begin{figure}[htbp]
\begin{center}
\epsfig{figure=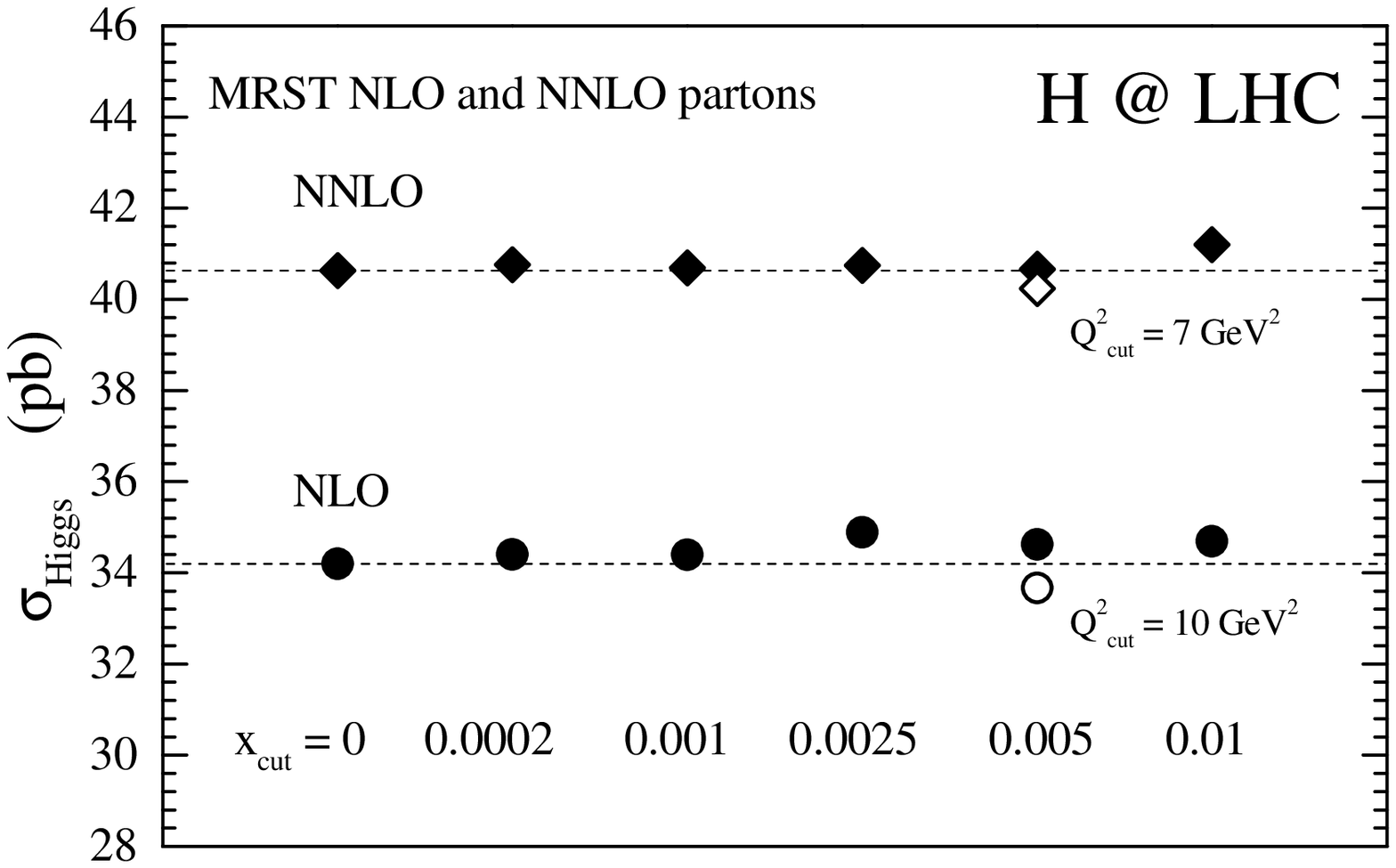,height=8cm}  
 \caption{\label{fig:HLHC} The same as Fig.~\ref{fig:HTev}, but for the LHC energy of $\sqrt s = 14~\TeV$.}
\end{center}
\end{figure}

\begin{figure}[htbp]
\begin{center}
\epsfig{figure=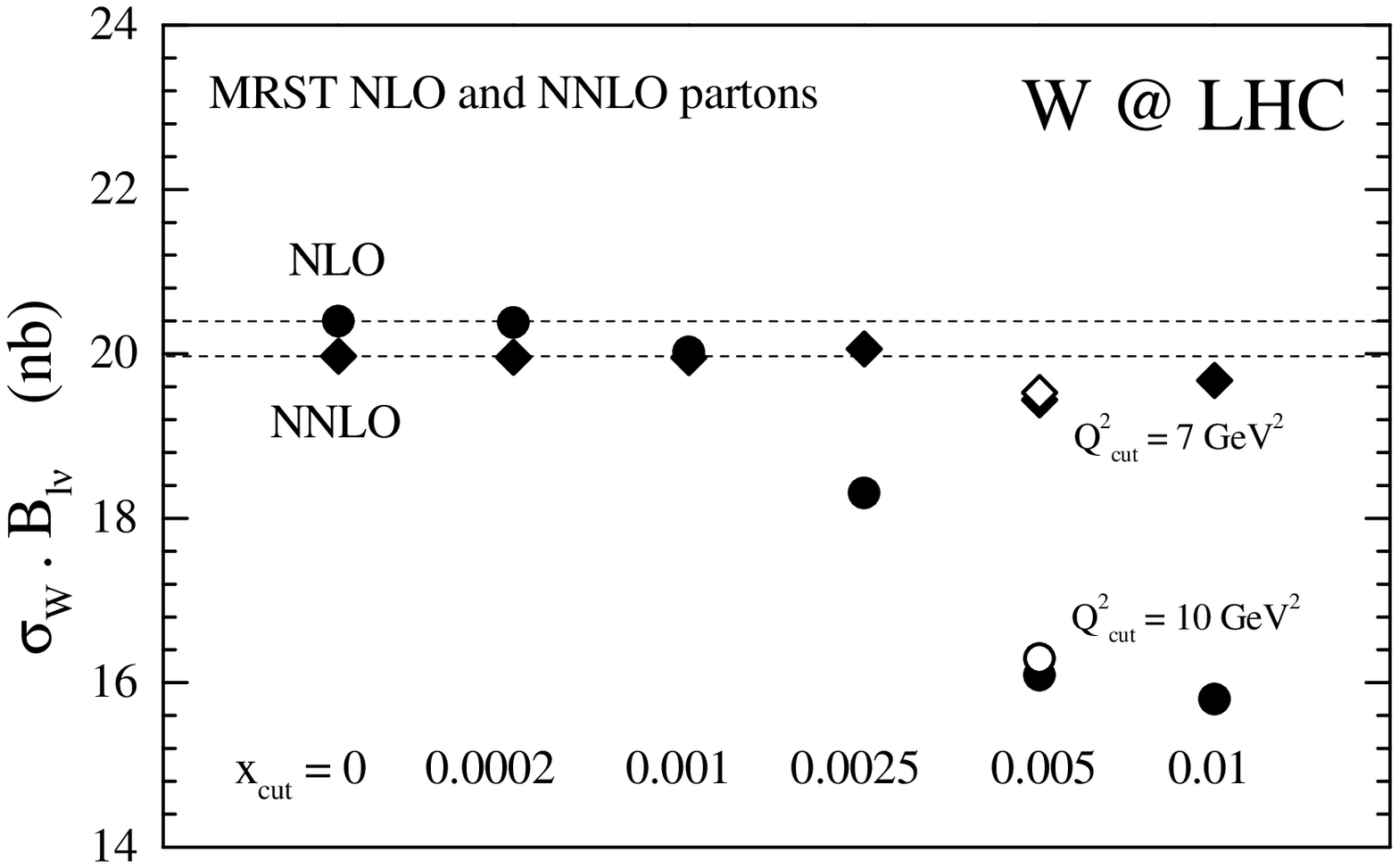,height=8cm}  
 \caption{\label{fig:WLHC} The same as Fig.~\ref{fig:WTev}, but for the LHC energy of $\sqrt s = 14~\TeV$.}
\end{center}
\end{figure}

Predictions for physical quantities are, like the value of $\alpha_S(M_Z^2)$, sensitive to
the `theoretical' uncertainties in the global parton analysis. For illustration, we show in Figs.~\ref{fig:HTev} --
\ref{fig:WLHC} the variation of the $W$ and Higgs cross section predictions for the Tevatron and LHC
 as a function of $x_{\rm cut}$ and $Q^2_{\rm cut}$.

The cross sections for $W$ and $H$ production at the Tevatron sample partons down to $x\simeq 0.005$, and so are
only directly sensitive to partons within the range of our most conservative cuts. However the cross sections can
still vary if we change the values of the cuts due to the readjustment of partons above $x_{\rm cut}$ and
$Q^2_{\rm cut}$, and for Higgs production, due to the different values of $\alpha_S(M_Z^2)$ extracted. This is
evident in Figs.~\ref{fig:HTev} and \ref{fig:WTev}.

The NLO prediction\footnote{For the Higgs cross section calculations described in this section we use
the full NLO QCD correction in the $m_t \gg M_H$ limit \cite{NLOHiggs} and the soft-virtual-collinear $x$--space
(SVC$_x$) approximation
to the full NNLO correction, again for $m_t \gg M_H$, taken from Ref.~\cite{SVC}. In both cases the factorisation
and renormalisation scales are set equal to $M_H$.}  for the cross section for the production of a Higgs of mass 120~GeV at the Tevatron rises
steadily as $x_{\rm cut}$ is increased, saturating at an increase of about 9\% for $x_{\rm cut} \simeq 0.005$.
This rise is due to the increase of the gluon distribution at moderate $x$ values as the value of $x_{\rm cut}$ is
raised, see Fig.~\ref{fig:f1}. There is a slight decrease in the value of $\alpha_S(M_Z^2)$ with increasing
$x_{\rm cut}$, but the effect of this on the Higgs cross section is completely outweighed by the large increase in
the gluon in the relevant $x$ range. We also see from Fig.~\ref{fig:HLHC} that, when $Q^2_{\rm cut}$ is increased to
$10~\GeV^2$, corresponding to our conservative set of NLO partons, the Higgs cross section is only increased by
1\%. This is mainly due to the drop in $\alpha_S(M_Z^2)$ to 0.116, but also due to a slight decrease in the gluon
for the relevant value of $x  \sim 0.06$ (for central rapidity production), compared to the case where only the cut
in $x$ is applied. This arises because the $Q^2$ cut of $10~\GeV^2$ eliminates the NMC $F^{p(d)}_2(x,Q^2)$ data
for $x\sim 0.05$, which prefer a steeper rise of $F_2(x,Q^2)$ with $Q^2$, and hence a larger gluon in this range
(as well as a larger value of $\alpha_S(M_Z^2)$). Nevertheless the value for the `conservative' set represents a
non-negligible increase in the Higgs cross section compared to the prediction of the default set.

From Fig.~\ref{fig:WTev} we also see that the NLO prediction for the cross section for $W$ production at the
Tevatron also rises steadily as $x_{\rm cut}$ is increased, saturating at an increase of about 2\% for $x_{\rm
cut}\simeq 0.005$. This is due to the increased evolution of the quarks driven by the increase of the gluon in the
relevant range. Again, when we also impose $Q^2_{\rm cut} = 10~\GeV^2$ the predicted value of the cross section
decreases. The increased value of $Q_{\rm cut}^2$ allows the input quarks to be larger for $x\simeq 0.05$ (too
large for the NMC data now cut out ), and the improvement in the quality of the fit requires less increase in $\partial
F_2/\partial \ln Q^2$. Compared to the quarks at $x\simeq 0.05$ obtained using $Q_{\rm cut}^2 = 2~\GeV^2$, the
quarks corresponding to $Q_{\rm cut}^2 = 10~\GeV^2$ cross in magnitude in the region of $Q^2 = 250~\GeV^2$, and
become 0.4\% smaller at $Q^2 = M_W^2$.

At NNLO we see from Figs.~\ref{fig:HTev} and \ref{fig:WTev} that the predictions for $W$ and $H$ 
production at the Tevatron are much
more stable to variations of $x_{\rm cut}$. The reason is that the increased evolution of quarks for $x\simeq
0.05$, due to the NNLO contributions to the splitting and coefficient functions, requires much less change in the
gluon, and consequently in  $\partial F_2/\partial\ln Q^2$ in this range. The additional imposition of $Q^2_{\rm
cut} = 7~\GeV^2$, however, reduces the Higgs cross section significantly, by about 5\%, 
even though at NNLO it leads to only a
slight decrease in the value of $\alpha_S(M_Z^2)$. This is because the loss of the NMC $F^{p(d)}_2(x,Q^2)$ data
for $x\sim 0.05$ below $Q^2$ of $10~\GeV^2$ has resulted in a large reduction in the gluon for $x\sim 0.1$, as
seen in Fig.~\ref{fig:NNLOcon}. The prediction for the $W$ cross section also falls, by about 1\%, for the same reason
as at NLO. Hence at NNLO the `conservative' parton set predicts a small decrease in both $\sigma_H$ and $\sigma_W$
compared to the prediction of the default parton set, while at NLO there is a small increase in both. As a
consequence, the `conservative' parton predictions show slightly greater convergence with increased perturbative order,
than the default predictions.

At the LHC, the cross sections for $W$ and $H$ production at central rapidity sample partons at $x =0.006$ and
$x=0.0085$ respectively, so these are safely predicted using the `conservative' partons. However, the {\em total}
cross sections sample a wide range of rapidity, and in fact probe down to $x\sim 0.00008$ for $W$ production and
$x\sim 0.0006$ for $H$ production, and hence the predictions using the `conservative' partons are not guaranteed to
be reliable. We present such predictions to exhibit where the reliability does break down, and to what extent.

We notice (Fig.~\ref{fig:HLHC}) that the predictions for the Higgs cross section at the LHC are actually rather stable to changes in the
value of $x_{\rm cut}$ at both NLO and NNLO. From Fig.~\ref{fig:fgluons} this is not too surprising. At the high
scales that are needed for Higgs production, the NNLO gluon is hardly changed compared to the default. The NLO
gluon, with $x_{\rm cut}=0.005$ applied, is a little larger in the central rapidity region, and only falls to a
much smaller value than the default at values of $x$ that are only making very small contributions to the total
cross section. Hence,  there is only a slight increase in the prediction for the Higgs cross section compared to
the default. When the additional cut in $Q^2$ is applied, in order to obtain the conservative sets, both the NLO
and NNLO predictions decrease slightly due to the decreases in $\alpha_S(M_Z^2)$. In fact, at NLO and NNLO the
predictions using the `conservative' sets finish very close to the default predictions (-1.5\% and -1\% for the changes
at NLO and NNLO respectively). This implies that there is
little theoretical error associated with Higgs production at the LHC due to the partons themselves. Indeed the
variation with $x_{\rm cut}$ and $Q^2_{\rm cut}$ at the LHC, exhibited in the figure, is evidently much smaller than the
correction in going from NLO to NNLO, that is, smaller than from the NNLO coefficient function contribution.\footnote{By way 
of calibration, the scale dependence of the $M_H = 120~\GeV$ Higgs cross section at the LHC is quoted as
$\pm 10\%$ (at NNLO) and $\pm 8\%$ (using NNLL soft-gluon resummation) in the recent study of Ref.~\cite{CFGN}.}

\begin{figure}[htbp]
\begin{center}
\epsfig{figure=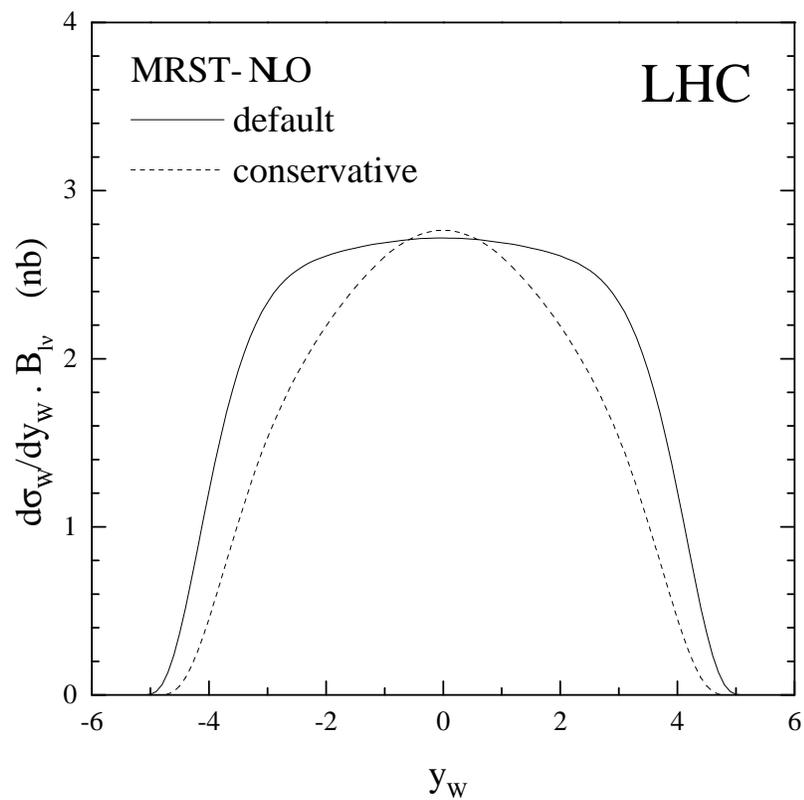,height=11cm}  
 \caption{\label{fig:wrap} The predictions for the rapidity distribution of the $W$ cross section at the LHC
for both the default MRST2002 set and for the `conservative' set.}
\end{center}
\end{figure}

For $W$ production at the LHC, Fig.~\ref{fig:WLHC}, the story is rather different. At NLO there is a steady, and rather dramatic
decrease in the predicted cross section as $x_{\rm cut}$ is increased. This culminates in a drop of $20\%$ for
$x_{\rm cut} =0.005$ (which is insensitive to the additional $Q^2$ cut). The reason for this is clear from
examination of Fig.~\ref{fig:f2}. For the sets with $x_{\rm cut}=0.005$ the quark distributions for $x<0.005$ are
much smaller than those of the default set. Also the gluon at small $x$ and low $Q^2$ is reduced, and therefore
the evolution of the quarks is slower than the default set. Hence, although at $Q^2 \sim 10^4~\GeV^2$ the quarks
at the central rapidity value of $x=0.006$ are actually much the same as in the default set, they quickly become
reduced at smaller $x$, and very much so for $x < 0.001$. However, much of the total cross section comes from $W$
rapidities corresponding to such low $x$ for one of the two quarks contributing to the reaction, and, as a
consequence, the contribution to the cross section at high rapidities is much reduced. This is illustrated in
Fig.~\ref{fig:wrap}, which shows the differential cross section as a function of the $W$ rapidity. At central rapidity,
the cross section is even increased slightly, but it falls away very quickly with increasing $|y_W|$, resulting in
the $20\%$ loss for the total. However, at NNLO there is far greater stability. There is a slight decrease in
predicted cross section with increasing $x_{\rm cut}$, which can be understood from the small decrease in the
quarks for $x$ in the region 0.0001--0.005, as compared to the default set, as seen in Fig.~\ref{fig:NNLOcon}.
However, as we have already commented for the gluon, the change in the very low $x$ partons with increasing
$x_{\rm cut}$ is very much reduced at NNLO, because the NNLO contributions themselves include important
corrections to the small-$x$ behaviour of the partons and structure functions.

We believe that the variation of the predicted cross sections with the value of $x_{\rm cut}$ and $Q^2_{\rm cut}$
gives a rough indication of the theoretical uncertainty due to the parton distributions. From 
the NNLO prediction for the cross
section for $W$ production at the Tevatron, we see from Fig.~\ref{fig:WTev} that, not only is the value stable, but
that the {\em theoretical} uncertainty is of order 1--2\%. This is similar in magnitude to the uncertainty due to
experimental errors, which was obtained in Ref.~\cite{MRST2002}. This gives a {\em total} error of about $\pm$2\%, 
which would imply that observing the $W$ production rate at the Tevatron (and comparing with the NNLO prediction) could
serve as a valuable luminosity monitor. The theoretical error,from the partons, for NNLO Higgs production 
at the Tevatron is rather
larger (about 5\%), due to the adjustment of the gluon over the whole $x$ range and changes in $\alpha_S$, when
the input assumptions to the fit are varied.

The variation at the LHC is relatively small for the Higgs cross section, which does not probe partons at too low
$x$, and is swamped by the higher-order corrections to the partonic cross section. However, it is potentially much
larger for the $W$ cross section at the LHC, which relies on lower $x$ partons. The $20\%$ variation in the $W$
cross section at the LHC at NLO implies that important theoretical corrections are required. The reduction of this
variation to $\sim 3\%$ at NNLO implies than much of this correction has occurred in going from NLO to NNLO.
However, in this case, the uncertainty is still larger than the change due to the NNLO contribution to the
partonic cross section and is the dominant theoretical uncertainty.\footnote{A very recent prediction of the $W$
and $Z$ cross sections at NLO and NNLO, with errors, has appeared in \cite{alekhinwz}. This disagrees with
our predictions by ${\cal O}(5\%)$, which is larger than the total quoted error in \cite{alekhinwz} ($2\%$ for the
Tevatron and $3\%$ for the LHC). The main reason for this discrepancy is almost certainly due to the absence of a
number of sets of data which determine the precise form of the quark distributions in the partonic
fit~\cite{alekhin03} that is used. Hence, in our spirit of determining parton distributions, we would deem this to
be a removable discrepancy.} The results in Section 4 imply that even at NNLO additional theoretical precision could be
obtained by further theoretical corrections, e.g. a correct inclusion of $\ln(1/x)$ terms at higher orders.
However, NNLO seems to be a great improvement on NLO, if one wishes to predict quantities sensitive to partons for
$x$ much lower than $0.005$.

\section{Conclusions} \label{sec:conclusions}

In a previous paper~\cite{MRST2002} we have already studied the uncertainties of parton distributions, and related
observables, arising from the errors on the experimental data used in the global parton analysis. However in that
paper we had already commented that in many cases the major uncertainty could be due to corrections to the
standard DGLAP evolution and due to assumptions used in the fit procedure, in other words due, collectively, to
so-called {\em theoretical} errors. In this paper we have studied a wide range of sources of theoretical error and
their potential consequences.

To begin, we investigated the possible corrections to standard fixed-order DGLAP analyses. The investigation was
performed in two alternative ways. First we made an empirical study in order to find those kinematic regions where
DGLAP evolution was fully consistent. This was done, at both NLO and NNLO, by gradually eliminating data until the
analyses were stable to further cuts. We found that this was best achieved by imposing separate cuts on $x$, $Q^2$
and $W^2$ of the data fitted.

\begin{itemize}
\item For $W^2$, stability was achieved by raising the cut from our default value of $12.5~\GeV^2$ to just
$15~\GeV^2$. However, in the region of low $W^2$, we noted that there exists some incompatibility between the data
sets.

\item For $x$, stability was achieved only for the relatively high value of $x_{\rm cut} = 0.005$. When this cut
was applied, the gluon distribution above this $x$ value increased, relative to the default set, at the expense of
the loss of the gluon at smaller $x$. This improved the quality of the fit to the Tevatron jet data, and also to
the structure function data for $x\sim0.01$ due to an increase in $\partial F_2/\partial\ln Q^2$. The $x_{\rm
cut}$ required was the same at NNLO as at NLO, but the modification of the gluon was considerably smaller at NNLO.

\item For $Q^2$, stability is reached for $Q_{\rm cut}^2 \sim 7$--10~$\GeV^2$ at both NLO and NNLO, the
convergence to stability being quite gradual as $Q_{\rm cut}^2$ is raised. The slow convergence indicates that
higher-order corrections, rather than higher-twist effects, are important at relatively low $Q^2$.
\end{itemize}

At both NLO and NNLO we also considered combinations of the above types of cuts. Thus we found the full kinematic
region where fixed-order DGLAP analysis is appropriate, together with the corresponding sets of conservative
partons. At NLO the domain is given by $W^2>15~\GeV^2$, $Q^2>10~\GeV^2$ and $x>0.005$, 
whereas at NNLO it is given by $W^2>15~\GeV^2$, $Q^2>7~\GeV^2$ and $x>0.005$.\footnote{In practice the 
conservative partons were obtained using $W^2_{\rm cut}=12.5~\GeV^2$, but raising $W^2_{\rm cut}$ to $15~\GeV^2$
has a negligible effect on the partons.} Within these regions these 
conservative sets of partons are most
reliable, but should not be used outside the domain.  Indeed, outside the region they may be completely
incompatible with the data. However, we note that, although the NLO and NNLO regions of stability are similar, the
NNLO partons are far closer to the default partons outside the stable domain, indicating smaller theoretical
uncertainty at NNLO.

Complementary to the above empirical study, we also made an explicit investigation of the following variety of
possible theoretical corrections.

\begin{itemize}
\item Higher-order parametric $\ln(1/x)$ corrections to the splitting functions were found to improve the quality of
the fit. This occurs both at small $x$, as expected, and also at larger $x$ where the partons are allowed to
re-adjust to a form similar to that of the conservative parton set (with $x_{\rm cut}>0.005$). The required
corrections were not the same at NNLO as at NLO, being overall a little smaller at NNLO, 
reflecting the important small $x$ contributions
introduced at NNLO.

\item Shadowing (or absorptive) corrections were introduced, but found to have little effect when the default
partons were used as the starting point. If the input gluon was forced to be even very slowly increasing at small
$x$, then evolution is too rapid even with shadowing corrections included. Further study would require the
simultaneous description of diffractive structure function data.

\item Parametric higher-twist contributions were introduced and quantified. These decreased as we progressed to
higher perturbative order, and, at NNLO, were only evident for $x\gtrsim0.5$. However, at both NLO and NNLO, the
change in the input partons is similar to that invoked by increasing $Q_{\rm cut}^2$. The high $x$, or
equivalently low $W^2$, higher-twist contribution is intrinsically related to the $\ln(1-x)$ resummation. This is
because the latter is inherently divergent and the ambiguity in the perturbation series must cancel that in the
higher-twist contribution. We found that the effect of the well-determined NNNLO high $x$ contribution was
important, and reduced the high $x$, higher-twist contribution still further. However this NNNLO contribution
demarks the perturbative order beyond which the series fails to converge. Hence we conclude that resummation
beyond NNNLO becomes indistinguishable from higher-twist corrections at high $x$.
\end{itemize}

As well as the theoretical corrections to the standard DGLAP evolution there are other potential sources of
uncertainty in the fitting procedure. Among these we considered the following.

\begin{itemize}
\item We found our input parameterization was sufficiently flexible to accommodate the data, and indeed there is a
certain redundancy evident. Hence we conclude that the form of our parameterization does not significantly
constrain the description, though a more efficient parameterization may be possible. Allowing our input gluon to be 
negative at small $x$ does have important consequences however.  

\item Heavy-target corrections are required when fitting to neutrino data. We found that, although our default
choice was not quite optimum, changes resulted in different data sets becoming more compatible with each other,
and led to only minimal changes in the partons.

\item Shadowing corrections are also required when fitting to deuterium structure functions. We found that there was 
a small amount of evidence for some high $x$ shadowing, and that the uncertainty in valence partons,
particularly $d_{\rm v}(x,Q^2)$ at high $x$, due
to the model uncertainty in deuterium shadowing is similar to the uncertainty due to the experimental 
errors on the data.  

\item The input strange quark sea distribution is parameterized as $\kappa(\bar u +\bar d)/2$, where the default
choice was $\kappa=0.5$. We found that both the global fit and the description of the CCFR dimuon data prefer a
smaller value of $\kappa\simeq 0.44$. This choice, which gives a smaller strange sea distribution and slightly
larger $\bar u, \bar d$, will be implemented in future fits. However it leads to only very small changes in most
physical quantities.

\item The possibility of isospin violations was investigated in both the sea and valence quark sectors. In the
former case, we found an that increase of $u_{\rm sea}^n$ (and a corresponding decrease of $d_{\rm sea}^n$), in
comparison to the default set, was preferred by the data at the 8\% level. In fact the acceptable range of
$u^n_{\rm sea}$ allowed an increase up to 18\% and a decrease down to 8\%. For the valence quarks there was no
significant improvement in the fit due to possible isospin violations, 
though a slight preference for $u^p_{\rm v}(x)>d^n_{\rm v}(x)$ and $d^p_{\rm v}(x)<u^n_{\rm v}(x)$
at high $x$ was noted. Changes in $u^n_{\rm v}$ of up to
$\pm10\%$ were permitted. From conservation of quark number the percentage violation for $d^n_{\rm v}$ is half
that for $u^n_{\rm v}$. For both valence and sea quarks the consequent change in the proton distributions in the
isospin-violating sets of partons is always 2\% or less. We observed that the possible isospin violation was 
certainly sufficient to account for the NuTeV $\sin^2\theta_W$ anomaly, and that the slight preference in the type
of violation for valence quarks is indeed in the correct direction to reduce this anomaly. 
We also discussed the possibility that $s\neq \bar s$,
noting that other studies have indicated a small inequality, the most recent of these again being 
in the correct direction to help 
resolve the NuTeV $\sin^2\theta_W$ anomaly.
\end{itemize}

We note that the most significant theoretical errors come from possible corrections to the fixed-order DGLAP
framework. Hence within the region of applicability, the conservative partons are the most reliable, and the
variation of partons, and of physical observables, under changes in $x_{\rm cut}$ and $Q^2_{\rm cut}$, give some
indication of the theoretical uncertainties, both inside and outside the conservative domain. The most reliable
extractions of $\alpha_S(M_Z^2)$ therefore follow from the conservative fits and are equal to
\be \alpha_S^{\rm NLO}(M_Z^2) \ = \ 0.1165 \pm 0.002({\rm expt}) \pm 0.003({\rm theory}), \ee
\be \alpha_S^{\rm NNLO}(M_Z^2) \ = \ 0.1153 \pm 0.002({\rm expt}) \pm 0.003({\rm theory}). \ee
In the NLO case this is considerably below our default determination, showing that the increase in $\alpha_S$ is
mimicking theoretical corrections beyond NLO DGLAP.

As additional tests of theoretical stability, we examined $W$ and Higgs cross sections at the Tevatron and the
LHC. At the Tevatron we only sample partons in the conservative domain. Nevertheless the NLO predictions vary
significantly as $x_{\rm cut}$ and $Q^2_{\rm cut}$ are changed due to the re-adjustment of the partons. The NNLO
predictions are much more stable, implying less theoretical uncertainty at NNLO. Indeed, the estimated 
total theoretical and
experimental uncertainty of about $\pm$2\% on $\sigma_W$ at the Tevatron offers an attractive and precise luminosity
monitor.

For non-central $W$ and Higgs production at the LHC, we probe partons below the conservative $x_{\rm cut}=0.005$.
For $\sigma_H$, which samples the gluon not too much lower than $x=0.005$, we still have reasonable stability,
especially at NNLO. On the other hand $\sigma_W$ samples quarks further below $x=0.005$ and the NLO prediction can
vary by up to 20\% with different choices of cuts. This implies that the theoretical uncertainty at small $x$ at
NLO is rather large. However, at NNLO, $\sigma_W$ is much more stable (varying by about 3\%), suggesting that the
theoretical uncertainty has been considerably reduced by the inclusion of NNLO splitting and coefficient
functions. We hope to confirm that this is still the case when the complete NNLO splitting functions become available.
Assuming that this is so, the total theoretical and experimental uncertainty at the LHC is about $\pm$4\%. Therefore
again it can serve as a good luminosity monitor. Work on resummations may be able to reduce the theoretical
uncertainty still further.

Therefore we conclude that theoretical uncertainties can be dominant in some kinematic regions, especially when
the physical quantity probes partons at small $x$ and/or small $Q^2$. There seems to be considerable advantage in
working at NNLO, as compared to NLO, but for real precision a few observables may have to await theoretical
developments in low $x$ and/or low $Q^2$ physics.

\section*{Acknowledgements}

We would like to thank Mandy Cooper-sarkar, Kevin McFarland and
Wu-Ki Tung for helpful conversations. RST would like to thank
the Royal Society for the award of a University Research Fellowship. RGR
would like to thank the Leverhulme Trust for the award of an Emeritus
Fellowship. The IPPP gratefully acknowledges financial support from the UK
Particle Physics and Astronomy Research Council.





\begin{thebibliography}{99}


\bibitem{Botje} 
M. Botje, Eur. Phys. J. {\bf C14} (2000) 285.

\bibitem{Giele}
 W.T. Giele and S. Keller, Phys. Rev. {\bf D58} (1998) 094023;\\
W.T. Giele, S. Keller and D.A. Kosower, {\tt hep-ph/0104052}.

\bibitem{Alekhin} 
S.I. Alekhin, Phys. Rev. {\bf D63} (2001) 094022.

\bibitem{CTEQLag} 
CTEQ Collaboration: D. Stump {\it et~al.}, Phys. Rev. {\bf D65} (2002) 014012.

\bibitem{CTEQHes} 
CTEQ Collaboration:
J. Pumplin {\it et~al.}, Phys. Rev. {\bf D65} (2002) 014013.

\bibitem{CTEQ6} 
CTEQ Collaboration: J. Pumplin {\it et~al.}, JHEP 0207:012 (2002).

\bibitem{H1Krakow} 
H1 Collaboration: C. Adloff {\it et~al.}, Eur. Phys. J. {\bf C21} (2001) 33.

\bibitem{H1new} 
H1 Collaboration: C. Adloff {\it et~al.}, {\tt hep-ex/0304003}.

\bibitem{ZEUSfit} 
A.M. Cooper-Sarkar, {\tt hep-ph/0205153}, J. Phys. {\bf G28} (2002) 2669;\\
ZEUS Collaboration: S. Chekanov {\it et~al.}, Phys. Rev. {\bf D67} (2003) 012007.

\bibitem{MRST2002}
  A.D.~Martin, R.G.~Roberts, W.J.~Stirling and R.S.~Thorne,  Eur. Phys. J. {\bf C28} (2003) 455.

\bibitem{HeavyFlav} 
M.A.G. Aivazis {\it et~al.}, Phys. Rev. {\bf D50} (1994) 3102;\\
J.C. Collins, Phys. Rev. {\bf D58} (1998) 094002;\\
R.S. Thorne and R.G. Roberts, Phys. Rev. {\bf D57} (1998) 1998, Phys. Lett. {\bf B421} (1998) 303, Eur. Phys. J. {\bf C19} (2001) 339;\\
W.-K. Tung, S. Kretzer and C. Schmidt, J. Phys. {\bf G28} (2002) 983.

\bibitem{H194} 
H1 Collaboration: C.~Adloff {\it et~al.}, Z. Phys. {\bf C72} (1996) 593.

\bibitem{BCDMS} 
BCDMS Collaboration: A.C.~Benvenuti {\it et~al.}, Phys. Lett. {\bf B223} (1989) 485.

\bibitem{E665} 
E665 Collaboration: M.R.~Adams {\it et~al.}, Phys. Rev. {\bf D54} (1996) 3006.

\bibitem{MRST2001}
A.D. Martin, R.G. Roberts, W.J. Stirling and R.S. Thorne, Eur. Phys. J. {\bf C23} (2002) 73.

\bibitem{MRSTtwist}
A.D. Martin, R.G. Roberts, W.J. Stirling and R.S. Thorne, Phys. Lett. {\bf B443} (1998) 301.

\bibitem{CF} 
E.B.~Zijlstra and W.L.~van~Neerven, Phys. Lett. {\bf
B272} (1991) 127; ibid. {\bf B273} (1991) 476; ibid. {\bf B297} (1992) 377; Nucl. Phys. {\bf B383} (1992) 525;\\
S. Moch and J.A.M.~Vermaseren, Nucl. Phys. {\bf B573} (2000) 853.

\bibitem{SF} 
S.A.~Larin, P.~Nogueira, T.~van~Ritbergen and
J.A.M.~Vermaseren, Nucl. Phys. {\bf B492} (1997) 338;\\
A.~Retey and J.A.M.~Vermaseren, Nucl. Phys. {\bf B604} (2001) 281;\\
S.~Moch, J.A.M.~Vermaseren and A.~Vogt, Nucl. Phys. {\bf B646} (2002) 181.

\bibitem{S47} 
S.~Catani and F.~Hautmann, Nucl. Phys. {\bf B427} (1994) 475; \\
V.S. Fadin and L.N. Lipatov, Phys. Lett. {\bf B429} (1998) 127; \\
G. Camici and M. Ciafaloni, Phys. Lett. {\bf B430} (1998) 349; \\
J. Bl\"{u}mlein and A. Vogt, Phys. Lett. {\bf B370} (1996) 149; \\
J.A. Gracey, Phys. Lett. {\bf B322} (1994) 141; \\
J.F. Bennett and J.A. Gracey, Nucl. Phys. {\bf B517} (1998) 241.

\bibitem{VV12} 
W.L. van Neerven and A. Vogt, Nucl. Phys. {\bf B568} (2000) 263; Nucl. Phys. {\bf B588} (2000) 345.

\bibitem{MRSTNNLO1} 
A.D. Martin, R.G. Roberts, W.J. Stirling and R.S. Thorne, Eur. Phys. J. {\bf C18} (2000) 117.

\bibitem{MRSTNNLO2} 
A.D. Martin, R.G. Roberts, W.J. Stirling and R.S. Thorne, Phys. Lett. {\bf B531} (2002) 216.

\bibitem{OWKID} 
N. Kidonakis and J.F. Owens, Phys. Rev. {\bf D63} (2001) 054019.

\bibitem{BFKL}      
L. N. Lipatov, Sov. J. Nucl. Phys. {\bf 23} (1976) 338;\\
E. A. Kuraev, L. N. Lipatov,  V. S. Fadin, Sov. Phys. JETP {\bf 45} (1977) 199;\\
I. I. Balitsky,  L. N. Lipatov, Sov. J. Nucl. Phys. {\bf 28} (1978) 338.

\bibitem{CH}        
S.~Catani and F.~Hautmann, Ref.~\cite{S47}.

\bibitem{BFKLNLL}   
V.S.~Fadin and L.N.~Lipatov, Phys. Lett. {\bf B429} (1998) 127.\\
G.~Camici and M.Ciafaloni, Phys. Lett. {\bf B430} (1998) 349.

\bibitem{RST}       
R.S.~Thorne, Phys. Lett. {\bf B474} (2000) 372; Phys. Rev. {\bf D64} (2001) 074005.

\bibitem{softglu} 
G. Sterman, Nucl. Phys. {\bf B281} (1987) 310;\\
    D. Appel, P. Mackenzie and G. Sterman, Nucl. Phys. {\bf B309} (1988) 259;\\
S. Catani and L. Trentadue, Nucl. Phys. {\bf B327} (1989) 323;\\
S. Catani, G. Marchesini and B.R. Webber, {\bf B349} (1991) 635.

\bibitem{VOGT} 
A. Vogt, Phys. Lett. {\bf B471} (1999) 97.

\bibitem{MQ} 
A.H.~Mueller and J.~Qiu, Nucl. Phys. {\bf B268} (1986) 427.

\bibitem{LR}
E.M.~Levin and M.G.~Ryskin, Phys. Rep. {\bf 189} (1990) 267.


\bibitem{KMRS}      
J.~Kwiecinski, A.D.~Martin, R.G.~Roberts and W.J.~Stirling, Phys. Rev. {\bf D42} (1990) 3645.

\bibitem{EHKQS}     
J.J.~Eskola, H.~Honkanen, V.J.~Kolhinen, J.~Qiu and C.A.~Salgado, Nucl. Phys. {\bf B660} (2003) 211.

\bibitem{RY}  
M.G.~Ryskin, plenary talk at XI International Workshop on Deep Inelastic Scattering, DIS03, St~Petersburg, April 2003.

\bibitem{AGK}  
V.~Abramovsky, V.N.~Gribov and O.V.~Kancheli, Sov. J. Nucl. Phys. {\bf 18} (1974) 308.

\bibitem{Bartels} 
J. Bartels and C. Bontus, Phys. Rev. {\bf D61} (2000) 034009.

\bibitem{TargetMass} 
H. Georgi and H.D. Politzer, Phys. Rev. {\bf D15} (1976) 1829.

\bibitem{renormalon} 
Yu.L. Dokshitzer and B.R. Webber, Phys. Lett. {\bf B352} (1995) 451.\\
M. Dasgupta and B.R. Webber, Phys. Lett. {\bf B382} (1996) 96.

\bibitem{DGE} 
E. Gardi, Nucl. Phys. {\bf B622} (2002) 365.

\bibitem{GR} 
E. Gardi and R.G. Roberts, Nucl. Phys. {\bf B653} (2003) 227.

\bibitem{MRSTkrakow} 
A.D. Martin, R.G. Roberts, W.J. Stirling and R.S. Thorne, Acta. Phys. Polon. {\bf B33} (2002) 2927.

\bibitem{CCFR} 
CCFR Collaboration: W.G.~Seligman {\it et~al.}, Phys. Rev. Lett. {\bf 79} (1997) 1213;\\
CCFR Collaboration: U.K.~Yang {\it et~al.}, Phys. Rev. Lett. {\bf 86} (2001) 2742.

\bibitem{DEUTSHAD} 
B. Badelek and J. Kwiecinski, Phys. Rev. {\bf D50} (1994) R4.

\bibitem{E139} 
J. Gomez {\it et al.}, Phys. Rev. {\bf D49} (1994) 4348.

\bibitem{Frankfurt} 
M. Frankfurt and M. Strikman, Phys. Rep. {\bf 160} (1988) 235.

\bibitem{bodekyang} 
U.K. Yang and A. Bodek, Phys. Rev. Lett. {\bf 82} (1999) 2467.

\bibitem{thomas} 
W. Melnitchouk {\it et al.}, Phys. Rev. Lett. {\bf 84} (2000) 5455.

\bibitem{akl} S.I. Alekhin, S.A. Kulagin and S. Luiti, {\tt hep-ph/0304210}.  

\bibitem{NuTeVF2} 
D. Naples {\it et al.}, {\tt hep-ex/0307005}.

\bibitem{CCFRmm} 
CCFR Collaboration: A.O.~Bazarko {\it et~al.}, Z. Phys. {\bf C65} (1995) 189.

\bibitem{NuTeV} 
M. Goncharov {\it et al.}, Phys. Rev. {\bf D64} (2001) 112006.

\bibitem{Rodionov}
E.N. Rodionov, A.W. Thomas and J.T. Londergan, Mod. Phys. Lett. {\bf A9} (194) 1799;\\
F.G. Cao and A.I. Signal, Phys. Rev. {\bf C62} (2000) 015203.

\bibitem{NuTevtheta} 
G.P. Zeller {\it et al.}, Phys. Rev. {\bf 88} (2002) 091802.

\bibitem{NuTeVtheta2} 
G.P. Zeller {\it et al.}, Phys. Rev. Lett. {\bf D65} (2002) 111103; Erratum {\it ibid} {\bf D67} (2003) 119902.

\bibitem{Olness} 
F. Olness, talk at XI International Workshop on Deep Inelastic Scattering,
DIS03, St~Petersburg, April 2003.

\bibitem{Wu-Ki} 
W.K. Tung, private communication, submitted for XXI International Symposium on Lepton and Photon
Interactions at High Energies, Fermilab National Accelerator Laboratory, Batavia, Illinois, USA, 11-16 August 2003.  

\bibitem{alekhin03}
S. Alekhin, Phys. Rev. {\bf D68} (2003)  014002; {\rm JHEP} {\bf 0302}, (2003) 015.

\bibitem{NLOHiggs} 
S.~Dawson, Nucl. Phys. {\bf B359} (1991) 283. \\
A.~Djouadi, M.~Spira and P.M.~Zerwas, Phys. Lett. {\bf B264} (1991) 440. 

\bibitem{SVC} 
S.~Catani, D.~de~Florian and  M.~Grazzini, {\rm JHEP} {\bf 0105}, (2001) 025. 

\bibitem{CFGN} 
S.~Catani, D.~de~Florian, M.~Grazzini and P.~Nason,  {\tt hep-ph/0306211}.

\bibitem{alekhinwz}
S. Alekhin, {\tt hep-ph/0307219}.

\bibitem{MV} 
A.~Milsztajn and M.~Virchaux, Phys. Lett. {\bf B274} (1992) 221.

\end{thebibliography}
\end{document}